\documentclass[preprint,12pt,authoryear]{elsarticle}

\usepackage{amssymb}
\usepackage{amsmath}
\usepackage[table]{xcolor}
\usepackage{graphicx}
\usepackage{tabularx}
\usepackage{xcolor}
\usepackage{float}
\usepackage{subcaption}
\usepackage{colortbl}
\usepackage{multirow}
\usepackage{array}

\usepackage{booktabs}
\usepackage{textgreek}
\usepackage{hyperref}
\usepackage[utf8]{inputenc}
\usepackage{bm}

\newcommand{\beginsupplement}{%
        \setcounter{table}{0}
        \renewcommand{\thetable}{S\arabic{table}}%
        \setcounter{figure}{0}
        \renewcommand{\thefigure}{S\arabic{figure}}%
     }

\journal{Medical Image Analysis}

\begin{document}

\begin{frontmatter}



\title{KongNet: A Multi-headed Deep Learning Model for Detection and Classification of Nuclei in Histopathology Images}

\author[label1,label2]{Jiaqi Lv}
\author[label1,label2]{Esha Sadia Nasir}
\author[label2]{Kesi Xu}
\author[label2]{Mostafa Jahanifar}
\author[label3]{Brinder Singh Chohan}
\author[label1,label2]{Behnaz Elhaminia}
\author[label1,label2]{Shan E Ahmed Raza}
\affiliation[label1]{organization={VSION Lab, Department of Computer Science, University of Warwick},
            city={Coventry},
            country={United Kingdom}}
\affiliation[label2]{organization={Tissue Image Analytics (TIA) Centre, Department of Computer Science, University of Warwick},
            city={Coventry},
            country={United Kingdom}}

\affiliation[label3]{organization={Department of Cellular Pathology, University Hospitals of Derby and Burton NHS Foundation Trust},
            country={United Kingdom}}

%% Abstract
\begin{abstract}
Accurate detection and classification of nuclei in histopathology images are critical for diagnostic and research applications. We present KongNet, a multi-headed deep learning architecture featuring a shared encoder and parallel, cell-type-specialised decoders. Through multi-task learning, each decoder jointly predicts nuclei centroids, segmentation masks, and contours, aided by Spatial and Channel Squeeze-and-Excitation (SCSE) attention modules and a composite loss function. We validate KongNet in three Grand Challenges. The proposed model achieved first place on track 1 and second place on track 2 during the MONKEY Challenge. Its lightweight variant (KongNet-Det) secured first place in the 2025 MIDOG Challenge. KongNet pre-trained on the MONKEY dataset and fine-tuned on the PUMA dataset ranked among the top three in the PUMA Challenge without further optimisation. Furthermore, KongNet established state-of-the-art performance on the publicly available PanNuke and CoNIC datasets. Our results demonstrate that the specialised multi-decoder design is highly effective for nuclei detection and classification across diverse tissue and stain types. The pre-trained model weights along with the inference code have been publicly released to support future research.
\end{abstract}

\begin{graphicalabstract}
\includegraphics[width=\linewidth]{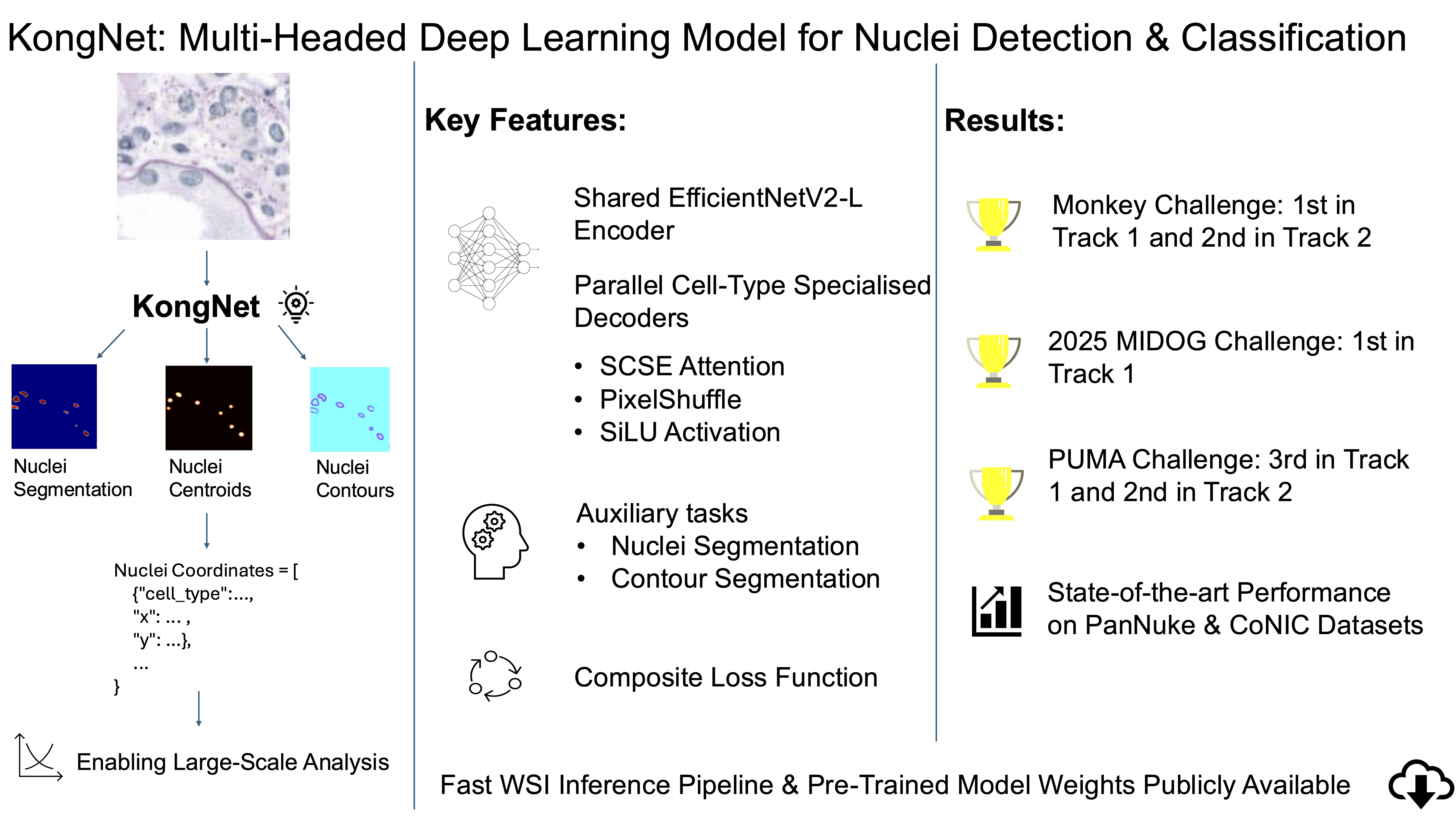}
\end{graphicalabstract}

\begin{highlights}
\item Multi-headed model with specialized decoders for nuclei detection.
\item Achieved 1st (Track 1) and 2nd (Track 2) in the MONKEY Challenge.
\item Lightweight variant secured 1st place in the 2025 MIDOG Challenge.
\item Shows state-of-the-art performance on the PanNuke and CoNIC datasets. 
\item KongNet secured top 3 rankings in the PUMA Challenge with no further optimization.
% \item Fast inference pipeline, code, and models are publicly released.
\end{highlights}

%% Keywords
\begin{keyword}
Nuclei Detection \sep Deep Learning \sep PAS-stained Images \sep H\&E-stained Images

\end{keyword}

\end{frontmatter}



\section{Introduction}

Accurate detection and classification of nuclei is a critical task in computational pathology. Pathologists routinely examine whole slide images (WSIs) of tissue specimens to assess morphological features that inform disease diagnosis and guide treatment planning. However, as these WSIs can contain millions of cells, manual quantification is time-consuming, tedious, and prone to subjectivity. 

The clinical relevance of automated analysis is particularly evident in oncology. For example, increasing attention is being given to the study of the Tumour microenvironment (TME), where interactions between tumour, immune, stromal and other cells types are believed to have a significant prognostic value~\cite{the_evolving_TME, TME_cell_cell_interactions, TME_cancer_progression}. Among the most promising biomarkers are tumour-infiltrating lymphocytes (TILs), which have been shown to correlate with clinical outcomes across various cancer types~\cite{SALTZ2018181,Brummel2023,El_Bairi2021-bc}. However, findings across studies are not always consistent; for instance, a higher percentage of TILs is associated with improved outcomes in triple-negative breast cancer, whereas the opposite correlation has been observed in HR-positive, HER2-negative subtypes~\cite{Valenza2023-hd}. These discrepancies underscore the need for accurate, generalisable, and scalable cell analysis tools to enable large-scale, systematic studies of the TME.

Beyond oncology, accurate quantification and subtyping of nuclei is also important in other domains. For example, in the histopathological evaluation of kidney transplant biopsies, the widely adopted Banff classification~\cite{2018Banff} defines 17 Banff Lesion Scores (BLS), ten of which involve quantifying the presence and spatial distribution of inflammatory cells across the renal compartments. Assessing these scores is a manual, labour-intensive process that is prone to both inter- and intra-observer variability, limiting consistency in transplant rejection grading. Moreover, a key component of the inflammatory response in transplant pathology is the infiltration of mononuclear leukocytes (MNLs), which comprise lymphocytes and monocytes, while these subtypes play different immunological roles, the Banff system does not differentiate them in routine scoring due to the difficulty of resolving their morphology in periodic acid–Schiff (PAS)-stained whole-slide images (WSIs). Automating the detection and classification of MNLs could significantly reduce pathologists' workload, improve reproducibility, and open the door to more granular immune phenotyping in kidney transplant diagnostics. However, this task is very challenging because the morphological features of lymphocytes and monocytes can appear to be very similar in PAS-stained images (see Figure~\ref{monkey_lymph_mono}). Although lymphocytes are generally smaller and rounder than monocytes, this distinction is often extremely subtle in PAS-stained images, as the cytoplasm of nuclei is unstained. This makes reliable differentiation a significant challenge. Pathologists usually have to rely on IHC staining to differentiate between them reliably. To accelerate progress in this direction, the MONKEY Challenge~\cite{ref_monkey_challenge} was launched as the first public benchmark for lymphocyte and monocyte detection in PAS-stained kidney WSIs. High-quality, cell-level annotations of lymphocytes and monocytes from multiple medical centres were released to encourage the development of algorithms capable of detecting and classifying MNLs. 

To address the above challenges, we introduce KongNet, a deep learning model that effectively captures nuanced nuclei features for accurate detection and classification.The name ``KongNet'' derives from our challenge team name, TIAKong, and has been retained for consistency with our challenge submissions and public model releases. While KongNet was originally developed to tackle the MONKEY challenge, which targeted at the detection and classification of MNLs in PAS-stained kidney biopsies, the core principles of KongNet's design are applicable to a broad range of tasks. KongNet has been evaluated across diverse benchmarks, including the PUMA Challenge~\cite{ref_puma_paper}, the 2025 MIDOG Challenge~\cite{MIDOG2025}, the PanNuke~\cite{pannuke_extension} and CoNIC~\cite{conic} datasets, achieving state-of-the-art (SOTA) results that extend far beyond its initial scope.

Our key contributions are summarised as follows:

\begin{enumerate}
    \item \textbf{A specialised multi-headed architecture:} We present a multi-headed deep learning model with specialised decoders and integrated attention modules for accurate detection and classification of nuclei. This architecture enables explicit class separation, reducing inter-class interference, and demonstrates superior performance in comparison to single-headed models.
    
    \item \textbf{Auxiliary objectives for robust feature learning:} We incorporate auxiliary tasks (nuclei and contour segmentation) to promote the learning of fine-grained morphological representations that benefit both detection and classification.
    
    \item \textbf{State-of-the-Art performance in Grand Challenges:} KongNet achieved \textbf{first place} in overall inflammatory cell detection and \textbf{second place} in subtype classification (lymphocytes and monocytes) in the MONKEY Challenge~\cite{ref_monkey_challenge}. It ranked \textbf{third} and \textbf{second} in track 1 and track 2 of the PUMA Challenge~\cite{ref_puma_paper} without further optimisation.

    \item \textbf{Lightweight model options:} We further develop and validate lightweight models including KongNet-SH and KongNet-Det for low-resource settings. We evaluated KongNet-Det during the 2025 MIDOG Challenge for mitosis detection, where it achieved \textbf{first place}. This demonstrates the architecture's adaptability to resource-constrained settings while maintaining SOTA detection quality.

    \item \textbf{Broad Generalisability across diverse datasets:} KongNet achieves SOTA results on the PanNuke~\cite{pannuke_extension} and CoNIC~\cite{conic} datasets, proving its effectiveness across multiple cancer types and staining protocols.
        
    \item \textbf{Public release of models:} We provide fast WSI inference pipeline and publicly release our model weights to support future research in computational pathology.
\end{enumerate}

\section{Related Works}
    
Object detection and classification have long been central challenges in computer vision, with many approaches being proposed in the literature. For example, the YOLO (You Only Look Once)~\cite{YOLO} family of models offers a fast, unified framework and has been widely adopted for natural image tasks. However, as reviewed by Debsarkar et al.~\cite{yolo_review}, YOLO’s performance in histopathology is limited by its dependence on large, well-separated objects and high-quality bounding box annotations, but these can be difficult to achieve in nuclei detection as the nuclei are small, dense, and sometimes overlapping. To address these challenges, researchers have developed specialised models. For example, Sirinukunwattana et al.~\cite{SCCNN} proposed a Spatially Constrained
Convolutional Neural Network (SC-CNN), where the model is trained to predict the probabilities of pixels being the centroids of nuclei, which means nuclei detection and classification could be done without having to perform segmentation. Raza et al.~\cite{mapde} proposed MapDe, which uses distance-based convolutional filters to emphasise nuclei centroid detection.  Sugimoto et al.~\cite{Sugimoto} introduced a multi-class detection model with a self-attention mechanism to improve separation of overlapping nuclei and classification of cancerous vs. non-cancerous cells. While effective, this method can be computationally expensive and challenging to scale up to larger number of cell classes. More generic architectures such as nnUnet~\cite{nnunet} and Faster-RCNN~\cite{faster-rcnn} have been successfully adapted for nuclei detection and classification of tumour cell in hereditary diffuse gastric cancer~\cite{Lomans2025}, and have shown good performance.

HoVer-Net~\cite{hovernet} pioneered a multi-task architecture using a shared encoder and three decoders, which are in turn responsible for binary segmentation, classification, and the prediction of horizontal and vertical gradient maps which are used to separate touching nuclei through Watershed. StarDist~\cite{stardist} predicts radial distances and per-pixel class probabilities using a multi-head U-Net, it achieved first place in the segmentation and classification track during the CoNIC Challenge~\cite{conic}. HoVer-NeXt~\cite{hovernext} attempts to improve HovVer-Net further by replacing HoVer-Net’s encoder with ConvNeXt and merging segmentation and gradient prediction tasks into a single decoder, with an optimised processing pipeline, HoVer-NeXt achieves a $17\times$ speed-up for inference on WSIs over HoVer-Net. DualU-Net~\cite{DualU-Net} attempts to simplify the HoVer-Net paradigm by using only two decoders and completely removing Watershed post-processing, while this showed improved computational efficiency, its classification accuracy did not improve significantly over HoVer-Net.

More recently, transformer-based approaches have also been explored. He et al.~\cite{TransNuSeg} proposed a multi-decoder transformer model for nuclei segmentation, contour prediction, and edge separation in clustered regions, but it does not support multi-class classification. CellViT~\cite{cellvit} showed strong performance on the PanNuke dataset by leveraging ViT encoders pre-trained on 104 million histology patches as well as the Segment Anything Model (SAM). NuLite~\cite{NuLite} similarly adopted transformer backbones but still relies on HoVer-Net-style pipelines, which involve the prediction of instance masks, classification maps, and distance/gradient maps. These predictions are combined during post-processing, which can often be slow and can require extra hyperparameter tuning. CellNuc-DETR~\cite{cellnucdetr}, a recent nuclei detection classification model using deformable detection transformer (Deformable-DETR) architecture with Swin Transformer encoder, shows SOTA performance on the PanNuke dataset, but its performance on other datasets has not been studied. 

InstanSeg~\cite{instanseg} took a different approach, using a multi-head U-Net to predict seed maps, positional, and conditional embeddings for nuclei segmentation. Although it achieves a strong segmentation performance on six public datasets, it does not support nuclei classification. A classifier could be trained to classify detected nuclei, but this would add additional complexity to the pipeline. 

In clinical workflows, pathologists often zoom in and out of WSIs to assess tissue context, which can be an important cue for cell classification. Recently, datasets incorporating both tissue and cellular annotations have emerged to study cell–tissue relationships. The OCELOT dataset~\cite{ocelot} demonstrated that incorporating tissue context can improve cell classification. However, the dataset consists of only two categories, tumour cells and non-tumour cells. The PUMA Challenge~\cite{ref_puma_paper} released a dataset with both tissue and nuclei annotations in H\&E-stained images of advanced melanoma. Torbati et al.~\cite{TeamLSM} developed a multi-stage pipeline to tackle this challenge. Their pipeline utilises tissue segmentation predictions to help refine nuclei classification. Although effective, their method requires five separate models and several manually defined ensemble rules, making it difficult to generalise to other domains. 

Although prior methods have advanced the field, they often suffer from a reliance on complex post-processing methods such as Watershed, or lack of support for classification. In addition, few methods have demonstrated strong adaptability across diverse staining protocols, tissue types, and cell types. In this study, we rigorously validated the performance of KongNet on five different histopathology datasets, covering multiple cell detection tasks and two distinct staining protocols.

\section{Method}

\subsection{Datasets}
The primary dataset used to develop KongNet is the MONKEY Challenge dataset~\cite{ref_monkey_challenge} (MONKEY dataset), which focuses on detecting monocytes and lymphocytes in PAS-stained kidney biopsies. To assess the generalisability of our approach across different tissue types, staining protocols, and cell types, we further train and evaluate the model on four additional datasets: the CoNIC dataset~\cite{conic}, PanNuke dataset~\cite{pannuke_extension}, the PUMA Challenge~\cite{ref_puma_paper} dataset (PUMA dataset), and the 2025 MIDOG challenge datasets. A summary of all the datasets used in work is listed in Table~\ref{datasets_table}.

\begin{table}
\centering
\caption{Summary of the datasets used in this study, including staining type, number of target classes (\# Classes), and number of annotated nuclei or mitotic figures (\# Nuclei). * indicates the number of mitotic figures sampled from the TCGA mitosis dataset for training, rather than the full number available in the original dataset.}
\label{datasets_table}
\small
\begin{tabularx}{\linewidth}{
>{\raggedright\arraybackslash}p{5.8cm}
>{\centering\arraybackslash}X
>{\centering\arraybackslash}X
>{\centering\arraybackslash}X
}
\toprule
\textbf{Dataset} & \textbf{Staining} & \textbf{\# Classes} & \textbf{\# Nuclei} \\
\midrule
MONKEY~\cite{ref_monkey_challenge} & PAS & 2 & 131,087 \\
CoNIC~\cite{conic} & H\&E & 6 & 535,063 \\
PanNuke~\cite{pannuke_extension} & H\&E & 5 & 186,744 \\
PUMA Track 1~\cite{ref_puma_paper} & H\&E & 3 & 97,429 \\
PUMA Track 2~\cite{ref_puma_paper} & H\&E & 10 & 97,429 \\
MIDOG++~\cite{MIDOG++} & H\&E & 1 & 11,937 \\
MITOS\_\allowbreak WSI\_\allowbreak CMC~\cite{CMC} & H\&E & 1 & 13,907 \\
MITOS\_\allowbreak WSI\_\allowbreak CCMCT~\cite{CCMCT} & H\&E & 1 & 44,880 \\
TCGA Mitosis~\cite{jahanifar2025mitosis} & H\&E & 1 & 161,739* \\
\bottomrule
\end{tabularx}
\end{table}

\subsubsection{MONKEY Dataset}

The MONKEY dataset was released for the MONKEY Challenge~\cite{ref_monkey_challenge} and consists of PAS-stained kidney biopsy WSIs for mononuclear leukocyte detection and classification. Expert pathologists annotated the centroids of two cell types, lymphocytes and monocytes, within 231 large regions of interest across 153 WSIs collected from six pathology departments (A--F).

The training set contains 81 cases from centres A--D. The preliminary test set contains 9 cases from centre E, while the final test set contains these same 9 cases together with an additional 19 cases from centre F. Both test sets were kept private by the organisers.

For each training case, a PAS slide scanned with the CPG profile using a P1000 WSI scanner (3DHistech) at Radboudumc is provided. For some cases, an additional diagnostic-profile PAS slide, a PAS slide scanned at the source institution, and an IHC slide stained with CD3 and CD20 antibodies are also available to support the annotation process. All WSIs are registered and have a resolution of 0.24 mpp. Since only CPG-profile WSIs were used during the challenge test phases, we also restricted model development to the CPG-profile slides.

From these WSIs, we extracted 14,315 patches of size $256\times256$ at 0.24 mpp, containing approximately 63,000 lymphocytes and 33,000 monocytes. We used 5-fold cross-validation, stratified by source institution to select the best model.

\subsubsection{CoNIC Dataset}

The CoNIC dataset~\cite{conic} was released for the 2022 CoNIC Challenge and consists of H\&E-stained images for nuclei detection and classification collected from 16 medical centres. The images are of size $256\times256$ pixels at $20\times$ objective magnification (approximately 0.5 mpp). The dataset contains approximately 535,000 annotated nuclei across six cell types: epithelial cells, plasma cells, lymphocytes, neutrophils, eosinophils, and connective cells.

We used the official training set for model development and performed 5-fold cross-validation to select the best model. Final benchmarking was performed on the official test set, which contains 103,150 annotated nuclei.

\subsubsection{PanNuke Dataset}

The PanNuke dataset~\cite{pannuke_extension} is a widely used benchmark for nuclei detection and classification in H\&E-stained images. It contains 186,744 annotated nuclei from 19 tissue types, extracted from over 20,000 WSIs, and covers five cell types: neoplastic cells, non-neoplastic epithelial cells, inflammatory cells, connective cells, and dead cells.

We trained and evaluated KongNet using the official 3-fold train--validation--test splits. The broad diversity of tissue types and nuclei morphology makes PanNuke a challenging benchmark for assessing model generalisation.

\subsubsection{PUMA Dataset}

The PUMA dataset was released for the PUMA Challenge~\cite{ref_puma_paper} and contains H\&E-stained images of advanced melanoma with both tissue-level and nuclei-level annotations. The full dataset comprises 155 primary melanoma ROIs and 155 metastatic melanoma ROIs across eight organ types. Each ROI is of size $1024\times1024$ pixels and was scanned at $40\times$ magnification (0.22 mpp).

The training set consists of 103 primary and 103 metastatic ROIs, with a total of 97,429 annotated nuclei. The preliminary test set contains 5 primary and 5 metastatic ROIs with 4,860 annotated nuclei, while the final test set contains the remaining 47 primary and 47 metastatic ROIs with 45,406 annotated nuclei. Both test sets were kept private by the organisers.

The metastatic samples span eight organs: skin, lymph node, soft tissue, brain, liver, lung, gastrointestinal tract, and bone. This makes the dataset effective for testing a model's capabilities to generalise to a wide range of organs.

The challenge includes two tracks. Track 1 focuses on three cell categories: tumour cells, tumour-infiltrating lymphocytes (TILs, including plasma cells), and all other cells. Track 2 focuses on 10 cell categories: tumour cells, lymphocytes, plasma cells, histiocytes, melanophages, neutrophils, stromal cells, epithelial cells, endothelial cells, and apoptotic cells. 

For model development, we extracted 3,280 patches of size $256\times256$ at $40\times$ magnification and used 5-fold cross-validation to select the best model.

\subsubsection{Datasets for Mitosis Detection (2025 MIDOG Challenge)}

To validate the lightweight KongNet-Det model, we evaluated it on mitosis detection datasets. The primary dataset, MIDOG$++$~\cite{MIDOG++}, was released for the 2025 MIDOG Challenge and is a single-class mitosis detection dataset containing 11,937 annotated mitotic figures across 503 tumour cases from seven tumour types.

For internal model development, we extracted patches of size $512\times512$ at approximately 0.25 mpp. We used MIDOG$++$ as the primary dataset for training, validation, and testing, we randomly selected 10\% of cases as a hold-out test set and performed 5-fold cross-validation on the remainder, stratified by tissue type.

To increase the diversity of the training data, we additionally sampled patches from three public mitosis datasets: MITOS\_WSI\_CMC~\cite{CMC}, which contains canine mammary carcinoma images and annotated 13,907 mitotic figures; MITOS\_WSI\_CCMCT~\cite{CCMCT}, which contains canine cutaneous mast cell tumour images and annotated 44,880 mitotic figures; and the Mitosis Dataset for TCGA Diagnostic Slides~\cite{jahanifar2025mitosis}, from which we sampled 161,739 automatically identified mitotic figures.

For external validation, we used the official test set from the 2025 MIDOG Challenge. The test set was curated to be particularly challenging in order to evaluate the robustness of the algorithms. This test set contains 120 cases from 12 tumour types, covering 12 distinct tumour types from both human and veterinary cases, including mitotic hotspots, random regions, and highly-challenging regions such as areas of inflammation and necrosis.

\subsection{Data Preprocessing}
\label{sec:Data Preprocessing}

\begin{figure}[t!]
\includegraphics[width=\textwidth]{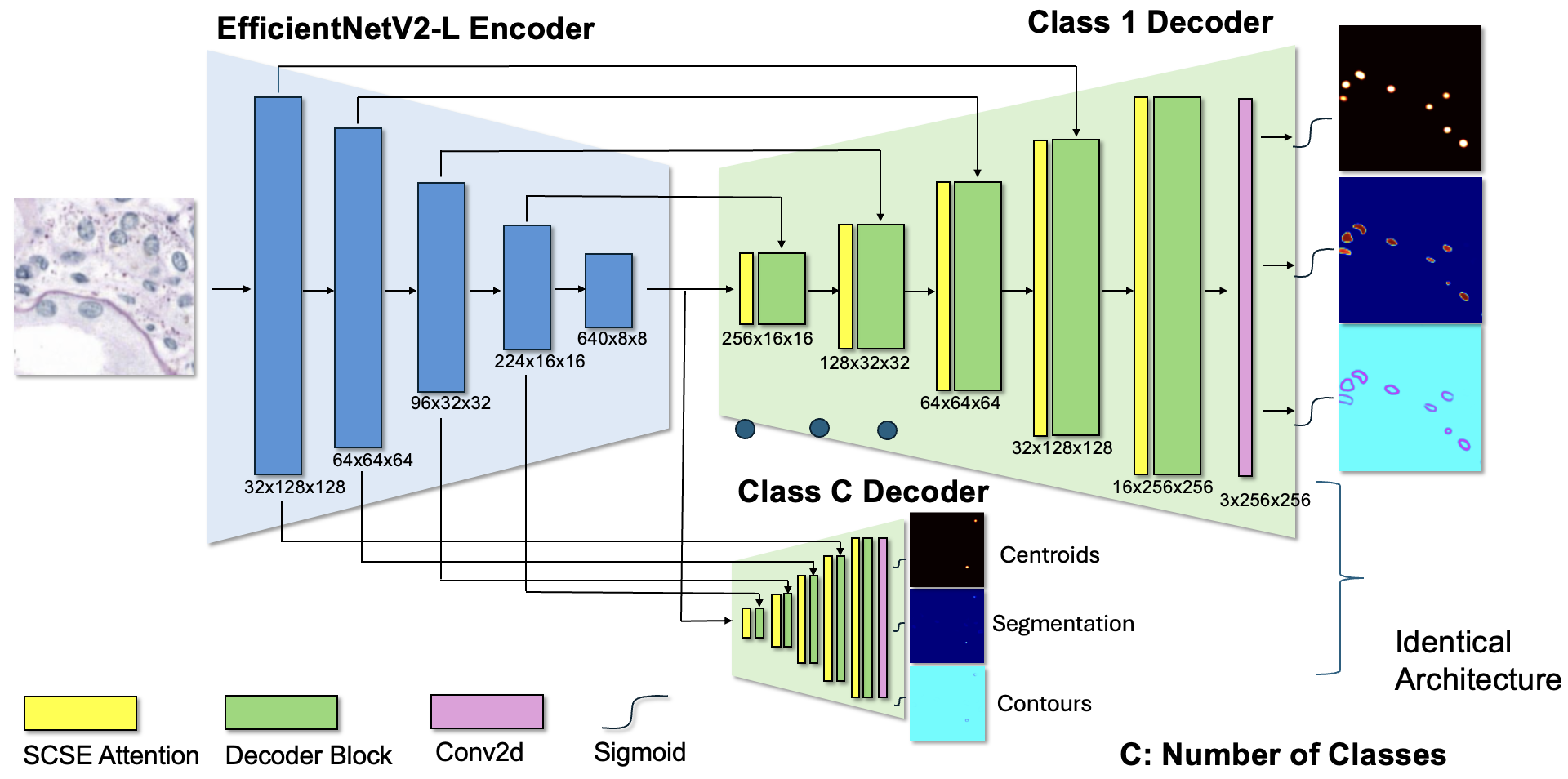}
\caption{Overview of the proposed architecture. The model consists of a shared EfficientNetV2-L encoder and ($C$) number of cell-type-specific decoders}. Each decoder learns discriminative morphological features for its target cell class by performing three tasks: nuclei centroid detection, nuclei segmentation, and nuclei contour segmentation.\label{model_overview_vis}
\end{figure}

\begin{figure}[t!]
\includegraphics[width=\textwidth]{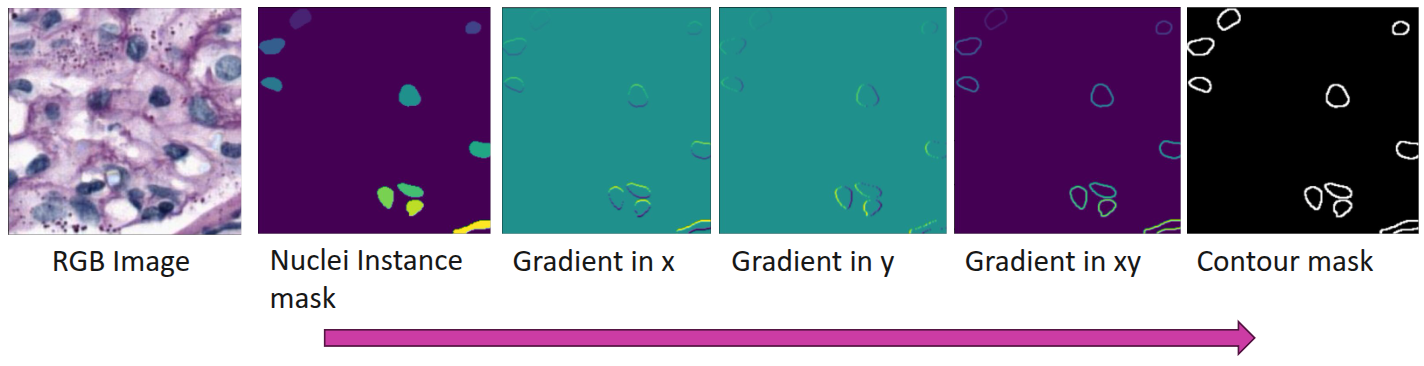}
\caption{Overview of the process of generating a nuclei contour mask from nuclei instance mask. Starting from the instance segmentation, Sobel filters are applied in the X and Y directions to obtain gradient maps. These maps are squared and summed to produce an edge map, which is then binarised to yield the final nuclei contour mask.} \label{contour_algo}
\end{figure}

An overview of the KongNet architecture is shown in Figure \ref{model_overview_vis}. KongNet utilises a multi-task framework for nuclei detection. In addition to its primary task of nuclei detection, we train KongNet to perform two auxiliary tasks: nuclei segmentation and nuclei contour segmentation. Therefore, we need to preprocess the data annotations to generate three corresponding ground truth masks for each image. 

For datasets with nuclei instance segmentation masks, we compute the centroid masks by finding the centre of mass for each nucleus instance. To generate contour masks, we employ a gradient based method. Sobel filter is applied in both the $X$ and $Y$ directions to the instance mask to obtain gradient maps. The sum of squared gradients produces an edge intensity map, which is then binarised to yield the final contour mask. This process is illustrated in Figure~\ref{contour_algo}.

For datasets providing only centroid (dot) annotations, such as the MONKEY dataset, we first generate the instance segmentation masks using NuClick~\cite{nuclick}, an interactive nuclei segmentation model pre-trained on the PanNuke dataset. The code and the model is publicly available through TIAToolbox~\cite{tiatoolbox}. Once the instance masks are obtained, contour masks are derived using the same gradient-based method described above.

Finally, to convert the centroid detection task into a semantic segmentation problem, we dilate nuclei centroids (dot annotations) using an elliptical structuring element. The diameter of the elliptical kernel is adjusted according to the image resolution (5 pixels at 0.5 mpp and 11 pixels at 0.25 mpp). These kernel sizes are selected heuristically based on the average size of human nuclei (around 10 $\mu$m in diameter~\cite{SUN2000184}), ensuring that the resulting disks are large enough for detection without overlapping neighbouring nuclei.

\subsection{Alleviating Class Imbalance}
To address the class imbalance problem commonly encountered in nuclei detection and classification tasks, we adapt and enhance the pixel-level sampling strategy used in HoVer-NeXt~\cite{hovernext_conic}. This technique assigns a higher sampling probability to training patches containing more cells of rare types, ensuring the model is exposed to a more balanced distribution of classes during training.

We introduce two key modifications to this approach. First, to make the strategy applicable to datasets without segmentation masks (such as MONKEY dataset), we use a mask-free method for estimating class prevalence. Instead of calculating pixel coverage from segmentation masks, we approximate the area occupied by each cell type by multiplying the number of annotated nuclei by a predefined average nucleus area (see Section \ref{sec:Data Preprocessing}). Second, to prevent extreme sampling probabilities for rare or common classes, we log-normalise the patch weights. Together, these enhancements create a robust weighted sampler that ensures the model is exposed to all cell types during training.

This sampling strategy can be described mathematically with the following equations. 
Let:
\begin{itemize}
    \item $P$ be the set of all patches in the training dataset.
    \item $C$ be the set of all cell classes, $C = \{c_0, c_1, ... ,c_n \}$, where $c_0$ is the background class.
    \item $A(p,c)$ be the area (in pixels) of class $c$ in a specific image patch $p$.
    \item $A_{patch}$ be the total area of a single patch.
\end{itemize}

First, the weight of each class $c$, denoted as $W_c$ is calculated based on its inverse frequency across the entire dataset (Eq.~\ref{sampler_eq_1}):
\begin{equation} \label{sampler_eq_1}
    \begin{aligned}
        W_c = \log{(\frac{\Sigma_{c' \in C} \Sigma_{p' \in P} A(p',c')}{\Sigma_{p' \in P} A(p', c)})}
    \end{aligned}
\end{equation}

Next, for any given patch $p$, its final sampling weight $W_p$, is the sum of the class weights $W_c$ weighted by the proportion of that patch occupied by each class $c$ (Eq.~\ref{sampler_eq_2}):
\begin{equation} \label{sampler_eq_2}
    \begin{aligned}
        W_p = \Sigma_{c \in C} (\frac{A(p,c)}{A_{patch}} \cdot W_c)
    \end{aligned}
\end{equation}

In essence, a class weight ($W_c$) is assigned to every cell type based on its inverse frequency across the entire dataset, meaning rare classes would receive a higher weight. Then, the final sampling weight of a training patch ($W_p$) is calculated as the sum of these class weights, scaled by the proportion of the area they occupy within the patch. Consequently, patches containing a larger percentage of rare cells are assigned a higher sampling probability.

\subsection{Model Architecture}
KongNet builds upon the established encoder-decoder paradigm commonly used in medical image analysis, but introduces a key architectural innovation: a multi-headed design with specialised decoders for each cell type. The core motivation is to reduce inter-class interference by allowing each decoder to focus on the unique morphological features of its target class, leading to improved detection and classification performance.

\subsubsection{Overall Architecture}
The model consists of a shared encoder and multiple parallel decoders. An overview of the model architecture is shown in Figure~\ref{model_overview_vis}.
\begin{itemize}
    \item \textbf{Encoder}: We use EfficientNetV2-L~\cite{ref_efficientnetv2_paper} initialised with ImageNet pre-trained weights. It processes the input image and extracts a hierarchy of multi-level features. The encoder is trained end-to-end with the rest of the network.
    \item \textbf{Decoders}:The extracted features are passed to multiple, parallel, cell-type-specific decoders. While the shared encoder learns features common to all nuclei, each decoder specialises in the distinct morphological characteristics of its assigned cell class. This design also simplifies post-processing. Since the primary objective of the model is to predict centroid maps, nuclei can be localised by detecting local maxima in the predicted centroid maps and then applying non-maximum suppression to resolve duplicate centroids. Consequently, KongNet does not require Watershed-based post-processing.
\end{itemize}

\subsubsection{Decoder Block Design}
\begin{figure}[t!]
\includegraphics[width=\textwidth]{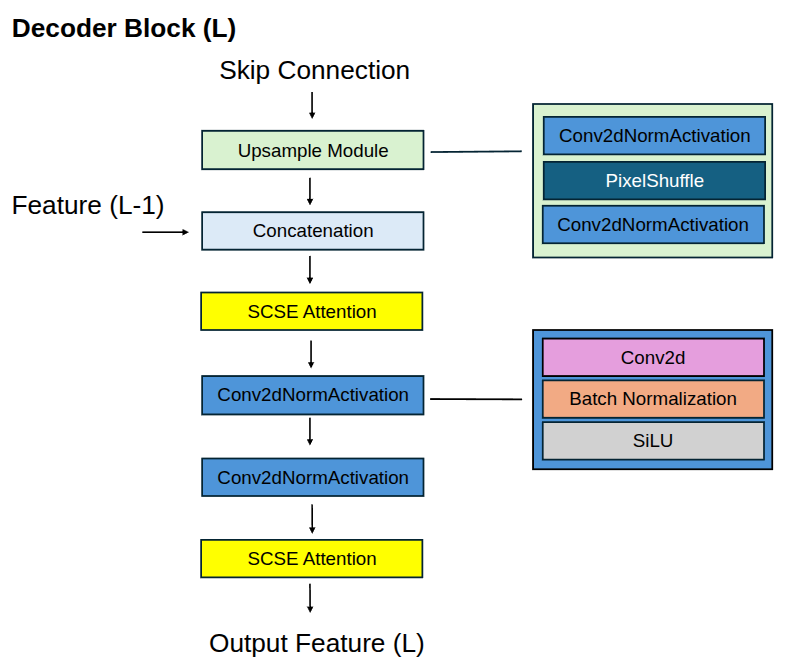}
\caption{Structure of a decoder block. Each block integrates Spatial and Channel Squeeze-and-Excitation (SCSE) modules to enhance feature recalibration. PixelShuffle is used for upsampling to better preserve spatial resolution and channel integrity. SiLU activation is applied to support smooth gradient flow and stable convergence.} \label{decoder_vis}
\end{figure}

Each decoder is composed of five decoder blocks which progressively upsample the features. Within each block, we incorporate several key components to enhance performance. A detailed illustration of the decoder block design is provided in Figure~\ref{decoder_vis}.
\begin{itemize}
    \item \textbf{SCSE Attention}: We integrate Spatial and Channel Squeeze-and-Excitation (SCSE) modules~\cite{scse} to apply channel-wise and spatial attention, recalibrating and refining the learned feature maps.
    \item \textbf{PixelShuffle Upsampling}: To better preserve both spatial and channel fidelity during upsampling, we use PixelShuffle~\cite{pixelshuffle}, an operation that rearranges elements from the channel dimension of a low-resolution feature map into spatial dimensions, increasing resolution without interpolation artifacts.
    \item \textbf{SiLU Activation}: The SiLU activation function is used throughout the decoder blocks to allow for a smooth gradient flow and stable training.
\end{itemize}

\subsubsection{Multi-Task Learning Framework}
\begin{figure}[t!]
\includegraphics[width=\textwidth]{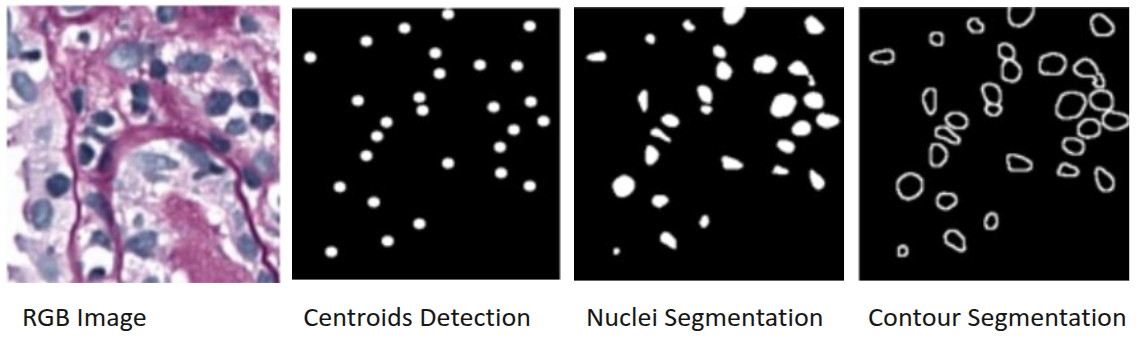}
\caption{An example from the MONKEY Dataset showing the three tasks performed by each decoder: nuclei centroid detection (primary task), nuclei segmentation (auxiliary task), and nuclei contour segmentation (auxiliary task).} \label{multitask_masks}
\end{figure}

To guide each decoder in learning robust morphological representations, we employ a multi-task learning framework. Specifically, in addition to nuclei detection, each decoder is trained using two auxiliary tasks for its specific cell type: nuclei segmentation and nuclei contour segmentation.
\begin{itemize}
\item \textbf{Nuclei Centroid Detection (primary task)}: A binary mask of dilated centroids.
\item \textbf{Nuclei Segmentation (auxiliary task)}: A binary mask of the full shape of the nuclei.
\item \textbf{Nuclei contour segmentation (auxiliary task)}: A binary mask of the nuclei boundaries.
\end{itemize}
A final sigmoid activation is applied to each of the three outputs, producing probability maps for each task, which are used to compute the loss during training. Figure~\ref{multitask_masks} shows example output masks for nuclei centroid detection, nuclei segmentation, and nuclei contour segmentation.

\subsubsection{Architecture Variants for Ablation Studies}
To validate the necessity of the independent, multi-decoder design, we introduce two additional KongNet variants for ablation studies:
\begin{itemize}
    \item \textbf{KongNet-SH(Single-Head)}: This model follows the identical encoder and decoder design as KongNet except that it uses a single, unified decoder which outputs prediction maps for all three tasks and all nuclei types. A diagram of the architecture is shown in Figure~\ref{KongNet-SH_vis}.
    \item \textbf{KongNet-Det(Detection-only)}: This lightweight model follows the identical encoder and decoder design as KongNet except that it uses a single decoder to predict only the nuclei centroids for each class (no auxiliary tasks). A diagram of the architecture is shown in Figure~\ref{KongNet-Det_vis}. We used this lightweight variant for the 2025 MIDOG Challenge, as it allowed for rapid development in a limited time frame without the need to pre-process mitosis annotations into nuclei segmentation or contour masks.
\end{itemize}
These variations allow us to explore the trade-offs between decoder parameter sharing, model complexity, and performance.

\subsection{Loss Functions}
To address the multi-class and multi-task nature of the problem, we designed a composite loss function $L$. We begin by defining the overall objective function and then detail its constituent parts.

\subsubsection{Overall Objective Function}
The overall loss function $L$ is the weighted sum of the individual class-specific losses $L_{Class\_k}$, regularised by a novel inter-class exclusion term, $L_{Interclass}$. The equation of the overall loss function is shown in Eq~\ref{overall_loss_eq}.

\begin{equation}\label{overall_loss_eq}
\begin{aligned}
    L &= \sum_{k=1}^C \lambda_k \cdot L_{Class\_k} + L_{Interclass}
    \end{aligned}
\end{equation}
Where $L_{Class\_k}$ is the loss for the cell class $k$, $\lambda_k \in {[0,1]}$ is its corresponding weight, and $C$ is the total number of classes excluding the background class.

This overall objective function has two key components:
\begin{enumerate}
    \item \textbf{Cell Class Loss Weighting ($\lambda_k$)}: The weights $\lambda_k$ are used for balancing the contributions of different cell classes. We employed two strategies for setting these weights:
    \begin{itemize}
    \item \textbf{Uncertainty-Based Weighting}: For the MONKEY and PUMA datasets, we used the method proposed by Kendall et al.~\cite{uncertainty_weighted_loss}, treating the $\lambda_k$ weights as learnable parameters that are dynamically optimised during training.
    \item \textbf{Fixed Equal Weighting}: For PanNuke and CoNIC datasets, we found that the dynamic approach suppressed the losses from rare classes. Therefore, we used a fixed, equal weighting scheme for all classes, ensuring a sufficient learning signal for under-represented cell types.
    \end{itemize}
    
    \item \textbf{Inter-class Exclusion Loss ($L_{interclass}$)}: To encourage mutually exclusive predictions, we introduce a global exclusion loss that penalises simultaneous activations across multiple classes for a given pixel. Unlike pairwise penalties, which can soften class boundaries, our term encourages a winner-takes-all behaviour. It is calculated as the product of the predicted probability maps across all classes. The equation is given in Eq.~\ref{interclass_exclusion_loss_eq}. 
    \begin{equation} \label{interclass_exclusion_loss_eq}
    \begin{aligned}
        L_{Interclass} &= \frac{1}{N} \sum_{i=1}^N\prod_{k=1}^C p_{k,i} \\
    \end{aligned}
    \end{equation}
    where $N$ is the total number of pixels, $C$ is the total number of cell classes. $p_{k,i} \in {[0,1]}$ is the predicted probability that pixel $i$ belongs to class $k$.
    
\end{enumerate}

\subsubsection{Class-Specific Loss}
$L_{Class\_k}$ represents the total loss for a single specialised decoder responsible for detecting cell type $k$. It is a combination of losses from the three simultaneous prediction tasks: centroid detection $L_{Centroid\_k}$, nuclei segmentation $L_{Seg\_k}$, and contour segmentation $L_{Contour\_k}$. We assign a smaller weight (0.5) to the contour loss, as it serves primarily as a guiding signal for the segmentation task. This is shown in Eq.~\ref{class_loss_eq}.
\begin{equation} \label{class_loss_eq}
\begin{aligned} 
    L_{Class\_k} = L_{Centroid\_k} + L_{Seg\_k} + 0.5 \cdot L_{Contour\_k}
\end{aligned}
\end{equation}

In the following equations, $g_{i,k} \in \{0,1\}$ denotes the binary ground truth for class $k$ at pixel $i$, and $p_{i,k} \in {[0,1]}$ is the corresponding predicted probability, $N$ is the total number of pixels in the image.

\begin{itemize}
    \item \textbf{Segmentation and Contour Loss}: For both auxiliary tasks: nuclei segmentation and contour segmentation, we use the same composite loss function $L_{BCE\_Dice}$, which combines Binary Cross-Entropy ($L_{BCE\_k}$) for pixel-level accuracy and Dice loss ($L_{Dice\_k}$) for capturing overall shape similarity. This is shown in Eq.~\ref{bce_dice_eq}.

    \begin{equation} \label{bce_dice_eq}
    \begin{aligned} 
    L_{BCE\_Dice} = L_{Contour\_k} = L_{BCE\_k} + L_{Dice\_k}
    \end{aligned}
    \end{equation}
    
    where $L_{BCE\_k}$ and $L_{Dice\_k}$ are defined as (Eq.~\ref{bce_eq}, Eq.~\ref{dice_eq}):

    \begin{equation} \label{bce_eq}
    \begin{aligned}
    L_{BCE\_k} &= -\frac{1}{N} \sum_{i=1}^{N} \left[ g_{i,k} \log(p_{i,k}) + (1-g_{i,k}) \log(1- p_{i,k}) \right]  \\
    \end{aligned}
    \end{equation}

    \begin{equation} \label{dice_eq}
    \begin{aligned}
    L_{Dice\_k} &= 1- Dice \\
                &= 1 - \frac{2 \sum_{i=1}^N p_{i,k} \cdot g_{i,k} + \epsilon}{\sum_{i=1}^N p_{i,k} + \sum_{i=1}^N g_{i,k} + \epsilon}
    \end{aligned}
    \end{equation}

    \item \textbf{Centroid Loss ($L_{Centroid\_k}$)}: For the primary task of nuclei centroid detection, we use a more robust combination of three losses: Jaccard Loss, Dice Loss, and Focal Loss. Dice and Jaccard losses focus on improving overall overlap, while Focal loss places more emphasis on hard-to-detect nuclei where the model often predicts with low confidence. This is shown in Eq.~\ref{centroid_loss}
    \begin{equation}\label{centroid_loss}
    \begin{aligned}
    L_{Centroid\_k} &= L_{Jaccard\_k} +L_{Dice\_k} + L_{Focal\_k} 
    \end{aligned}
    \end{equation}
    
    where $L_{Jaccard\_k}$ and $L_{Focal\_k}$ are defined as (Eq.~\ref{jaccard_eq}, Eq.~\ref{focal_eq}):
    \begin{equation}\label{jaccard_eq}
    \begin{aligned}
       L_{Jaccard\_k} &= 1- Jaccard\_k \\
            &= 1 - \frac{\sum_i p_{i,k} \cdot g_{i,k} + \epsilon}{\sum_{i,k} p_{i,k}^2 + \sum_{i} g_{i,k}^2 -  \sum_{i} p_{i,k} \cdot g_{i,k} + \epsilon}  \\
    \end{aligned}
    \end{equation}

    \begin{equation} \label{focal_eq}
    \begin{aligned}
    &L_{Focal\_k} = \\
    &-\frac{1}{N} \sum_{i=1}^{N} \left[ \alpha g_{i,k} (1-p_{i,k})^\gamma \log(p_{i,k}) + (1-\alpha)(1-g_{i,k}) p_{i,k}^\gamma \log(1-p_{i,k}) \right] \\
    \end{aligned}
    \end{equation}
    $\alpha = 0.25$ and $\gamma = 2$.
\end{itemize}

\subsection{Evaluation Metrics}
To ensure consistency and enable direct comparison with other state-of-the-art methods, we adopted the official evaluation protocols for each dataset and Grand Challenge.

\subsubsection{MONKEY Challenge Evaluation}
The primary metric for the MONKEY Challenge is the Free Response Operating Characteristic (FROC) analysis. In FROC analysis, the true positive rate (TPR) is plotted against the average number of false positives (FP) per $mm^2$ across all slides and across a range of detection confidence thresholds. A predicted detection is considered a true positive (TP) if its centroid lies within a predefined distance margin from a ground truth annotation. For lymphocytes and monocytes, the distance margins are set to $4\mu m$ and $5\mu m$, respectively. For combined inflammatory cell detection, a margin of $5\mu m$ is used.

Using these definitions, true positives (TP), false positives (FP), and false negatives (FN) are computed for each model to generate the corresponding FROC curve. The final FROC score is calculated as the average sensitivity measured at five pre-specified false positive rates (FP$/mm^2$), as defined by the challenge organisers.

It is important to note that inflammatory cells, monocytes, and lymphocytes are evaluated independently. This allows for a soft classification of cells, where a single cell can be assigned to multiple classes with different probabilities.

\subsubsection{Evaluation of PanNuke Models}
For the PanNuke dataset, we followed the official evaluation protocol, which uses a combined detection and classification F1 score. A prediction is matched to a ground truth nucleus if their centroids are within a 12-pixel radius. The evaluation penalises both detection errors and classification errors within the F1 score calculation.

The overall detection F1 score, precision, and recall ($F_{1d}$, $P_d$, $R_d$) are calculated as:
\begin{align} \label{pannuke_detection_f1_eq}
    F_{1d} &= \frac{2TP_d}{2TP_d + FP_d + FN_d}, \\
    P_d &= \frac{TP_d}{TP_d + FP_d}, \\
    R_d &= \frac{TP_d}{TP_d + FN_d},
\end{align}

Following detection matching, the classification F1 score, precision, and recall per class $c$ ($F_{1c}$, $P_c$, $R_c$) are computed as:
\begin{align} \label{pannuke_type_f1_eq}
    F_{1c} &= \frac{2(TP_c + TN_c)}{2(TP_c + TN_c) + 2FP_c + 2FN_c + FP_d + FN_d}, \\
    P_c &= \frac{TP_c + TN_c}{TP_c + TN_c + 2FP_c + FP_d}, \\
    R_c &= \frac{TP_c + TN_c}{TP_c + TN_c + 2FN_c + FN_d},
\end{align}

\subsubsection{Evaluation for CoNIC, PUMA and MIDOG Models}
For CoNIC data, PUMA and MIDOG challenges, detection quality is computed using the F1 score for each cell class, defined as:
\begin{align} \label{f1_detection_quality_eq}
F1_{t} &= \frac{2 \cdot TP_t}{2 \cdot TP_t + FP_t + FN_t},
\end{align}
where $TP_t$, $FP_t$, and $FN_t$ denote the true positive, false positive, and false negative counts for class $t$, respectively.

For \textbf{CoNIC dataset}, the $TP_t$, $FP_t$, and $FN_t$ counts are first aggregated across all test images for each cell class before the F1 score is calculated. While the official CoNIC metric is based on segmentation intersection-over-union (IoU), our model is designed to predict centroids. Therefore, consistent with other detection-based methods, we match the predicted and ground truth centroids using a fixed radius of 6 pixels (the standard matching radius for $0.5mpp$ resolution as described in the CellViT paper~\cite{cellvit}). 

The \textbf{PUMA Challenge} also uses the per-class F1 score as its primary metric. However, unlike CoNIC, the F1 score is calculated on a per-image basis for each class, and these scores are then averaged across all images to produce the final result. Predicted and ground truth centroids are matched using a fixed radius of 15 pixels.

The evaluation of the \textbf{2025 MIDOG Challenge} is similar to CoNIC, where the performance is calculated by aggregating counts across all images before computing the final F1 score. A predicted mitosis is matched to a ground truth annotation if their centroids are within a radius of $7.5\mu m$.

\section{Results}
For consistency with the original sources, results from prior publications and official challenge leaderboards are reported using the values and decimal precision provided in those sources. Consequently, the number of decimal places is not always uniform across datasets and benchmarks.

\subsection{MONKEY Challenge Results}
As the model was initially developed for the MONKEY challenge, we first report the performance of the model on our internal cross validation for the MONKEY Challenge. The MONKEY Challenge comprised two tracks: Track 1, ranked by the FROC score for overall inflammatory cell detection, and Track 2, ranked by the average FROC score for lymphocyte and monocyte classification.

\subsubsection{Internal Validation Confirms Multi-Decoder Advantage}
We first performed a rigorous five-fold cross-validation on the MONKEY training set to optimise our architecture, the results are reported in Table~\ref{monkey_table_local_validation}. These results clearly established the superiority of our proposed multi-decoder design. The baseline KongNet-Det (detection-only) showed the lowest performance, while KongNet-SH (single-head multi-task) offered improvement. However, the full KongNet, with its independent, class-specific decoders, achieved the best results.

\begin{figure}[t!]
    \centering
    \includegraphics[width=\linewidth]{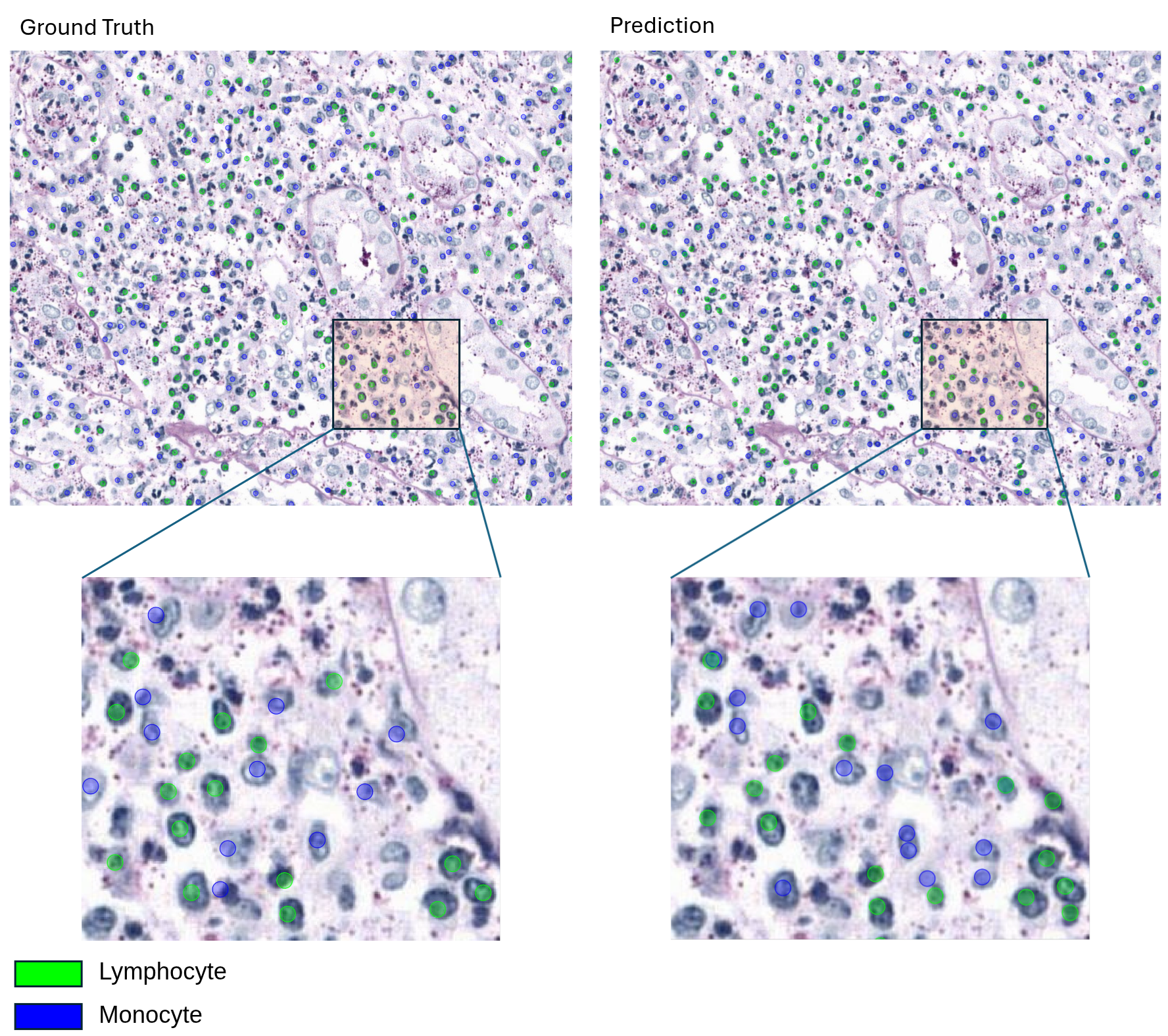}
    \caption{Example image from the MONKEY dataset showing ground truth annotations for lymphocytes (green) and monocytes (blue), alongside KongNet predictions.}
    \label{monkey_pred_vis}
\end{figure}

Based on this, we created a final, wider version, KongNet (Wide), by increasing the channel capacity in the decoder blocks to $(512, 256, 128, 64, 32)$. This configuration demonstrated the highest performance in our internal validation and was selected as our submission for the official challenge. A qualitative example of its predictions is shown in Figure~\ref{monkey_pred_vis}.

\subsubsection{Preliminary Leaderboard Performance}
On the preliminary leaderboard, KongNet (Wide) ranked a strong second overall, achieving competitive FROC scores that were close to the top-performing team and demonstrated a clear advantage over other participants (See Table~\ref{monkey_table_preliminary}). The preliminary test set consisted of only nine cases from a single unseen medical centre. To avoid overfitting to this small sample, we deliberately refrained from further fine-tuning our model or post-processing parameters based on these results.

\subsubsection{Final Leaderboard Performance}
On the final leaderboard, KongNet (Wide) secured first place in overall inflammatory cell detection (FROC: 0.3930) and first place in lymphocyte detection (FROC: 0.4624), outperforming all other teams in these categories (Table~\ref{monkey_table_final}). For the monocyte class, our model ranked second (0.2392), behind Team InstanSeg who leveraged over two million synthetically generated annotations from IHC images~\cite{instanseg_monkey}. This may have contributed to their superior detection performance for monocytes, an under-represented class in the original dataset.

Our model's performance on lymphocyte and monocyte detection increased significantly from the preliminary to the final phase, despite the final test set introducing a domain shift with images from two previously unseen medical centres. This demonstrates robustness and generalisation capability of the KongNet architecture compared to other SOTA models such as YOLO, Faster-RCNN, and CellViT. These results validate the effectiveness of using specialised decoders for the highly challenging task of differentiating mononuclear leukocytes in PAS-stained images.

\begin{table}[t!]
\centering
\caption{Summary of detection results (FROC) in the MONKEY Challenge \textbf{Final Test Leaderboard}. $^1$InstanSeg + Classifier~\cite{instanseg_monkey}. $^2$DETR and YOLOv5-L~\cite{AIRA_Monkey}. $^3$YOLOv11 + Classifier~\cite{st_medical_monkey}. $^4$Customised CNN~\cite{biototem_monkey}. $^5$Two Faster R-CNN models~\cite{ouradiology_monkey}. $^6$CellViT~\cite{zip_lab_unimore_monkey}. $^7$Faster R-CNN trained by organisers of the MONKEY Challenge~\cite{ref_monkey_challenge}.}
\label{monkey_table_final}
\begin{tabularx}{\columnwidth}{
    >{\raggedright\arraybackslash}p{5cm}
    >{\centering\arraybackslash}X
    >{\centering\arraybackslash}X
    >{\centering\arraybackslash}X
}
\toprule
Method & Inflammatory Cells $\downarrow$ & Lymphocytes & Monocytes \\
\midrule
KongNet (Wide) & \textbf{0.3930} & \textbf{0.4624} & 0.2392 \\
Team InstanSeg$^1$ & 0.3875 & 0.4515 & \textbf{0.2626} \\
Team AIRA Matrix$^2$ & 0.3517 & 0.4471 & 0.1906 \\
Team ST Medical$^3$ & 0.3316 & 0.3935 & 0.1268 \\
Team BioTotem$^4$ & 0.3130 & 0.3702 & 0.1230 \\
Team ouradiology$^5$ & 0.2824 & 0.3703 & 0.1550 \\
Team Zip Lab UNIMORE$^6$ & 0.2470 & 0.3049 & 0.0699 \\
Baseline$^7$ & 0.2282 & 0.3120 & 0.1220 \\
\bottomrule
\end{tabularx}
\end{table}

\subsection{CoNIC Results}
To test KongNet's multi-class classification capabilities on a diverse range of cells, we evaluated it on the challenging CoNIC dataset, which comprises six morphologically diverse cell types: neutrophils, epithelial cells, lymphocytes, plasma cells, eosinophils, and connective tissue cells. 

\begin{figure}[t!]
    \centering
    \includegraphics[width=\linewidth]{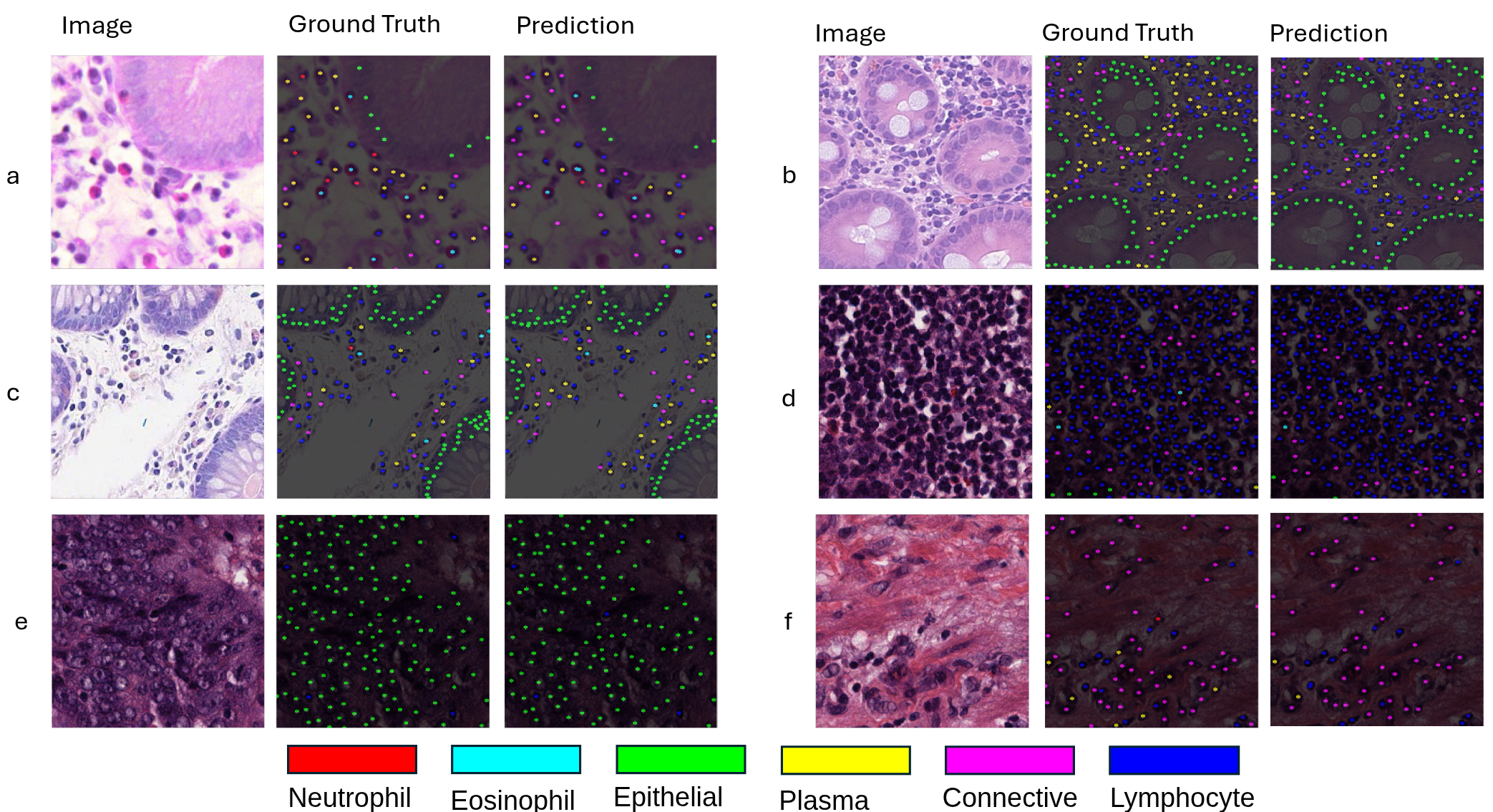}
    \caption{Examples from the CoNIC test set showing the input image, ground truth annotations, and KongNet predictions.}
    \label{conic_pred_vis}
\end{figure}

We compare KongNet against the top published submissions from the CoNIC Challenge. The results are shown in Table~\ref{conic_table}. KongNet achieved a state-of-the-art class-average F1 score of 0.653, outperforming all other published methods, including the winner of the original challenge (StarDist). KongNet secured the highest F1 score in five of the six cell categories, with StarDist showing a small advantage for plasma cells (0.612 vs. 0.596). The qualitative results in Figure~\ref{conic_pred_vis} further demonstrate KongNet's robust detection quality across images with significant variations in staining, lighting, and cell morphology.

This SOTA performance was achieved using the same architecture optimised for the PAS-stained MONKEY dataset. This successful transfer from a two-class renal pathology task to a six-class oncology benchmark strongly underscores the inherent generalisability of our specialised multi-decoder approach, highlighting its effectiveness across different tissue types, cell morphologies, and staining protocols.

\begin{table}[t!]
\centering
\caption{Summary of detection results (F1) on the CoNIC Test Dataset for each cell type and the average F1 across cell types. *Ground truth and predicted coordinates were matched using centroids with a radius of 6~pixels. Neu: Neutrophil; Epi: Epithelial; Lym: Lymphocyte; Pla: Plasma; Eos: Eosinophil; Con: Connective. Class Average: Average of the F1 scores across cell types.}
\label{conic_table}
\begin{tabularx}{\textwidth}{
>{\raggedright\arraybackslash}p{3.8cm}
>{\centering\arraybackslash}p{0.8cm}
>{\centering\arraybackslash}p{0.8cm}
>{\centering\arraybackslash}p{0.8cm}
>{\centering\arraybackslash}p{0.8cm}
>{\centering\arraybackslash}p{0.8cm}
>{\centering\arraybackslash}p{0.8cm}
>{\centering\arraybackslash}X
}
\toprule
Method & Neu & Epi & Lym & Pla & Eos & Con & Class Average $\downarrow$ \\
\midrule
KongNet* & \textbf{0.510} & \textbf{0.818} & \textbf{0.707} & 0.596 & \textbf{0.591} & \textbf{0.695} & \textbf{0.653} \\
EPFL (StarDist)~\cite{stardist} & 0.476 & 0.762 & 0.696 & \textbf{0.612} & 0.497 & 0.653 & 0.616 \\
MDC Berlin~\cite{hovernext_conic} & 0.380 & 0.739 & 0.676 & 0.578 & 0.534 & 0.644 & 0.592 \\
Pathology AI~\cite{Pathology-AI-CoNIC} & 0.415 & 0.741 & 0.693 & 0.562 & 0.445 & 0.627 & 0.581 \\
Arontier~\cite{Arontier_CoNIC} & 0.343 & 0.726 & 0.651 & 0.583 & 0.472 & 0.633 & 0.568 \\
AI\_Medical~\cite{AI_Medical_CoNIC} & 0.377 & 0.735 & 0.675 & 0.587 & 0.450 & 0.632 & 0.513 \\
\bottomrule
\end{tabularx}
\end{table}

\subsection{PanNuke Results}
We tested KongNet on the PanNuke dataset, a highly diverse benchmark featuring 19 tissue types. Using the official evaluation protocol, we report the average F1 scores in Table~\ref{pannuke_table}. 

KongNet achieved the highest class-average F1 score (0.674), demonstrating a more balanced performance profile than other leading methods. Specifically, our model excelled in classifying cell types often identified by their distinct individual morphology: inflammatory (0.72), connective (0.70), and dead cells (0.59). In contrast, the transformer-based CellNuc-DETR performed best on neoplastic and epithelial cells, categories where broader tissue context may play a larger role. This suggests a complementary strength of our CNN-based, multi-decoder approach in capturing fine-grained local features.

\begin{figure}[t!]
    \centering
    \includegraphics[width=\linewidth]{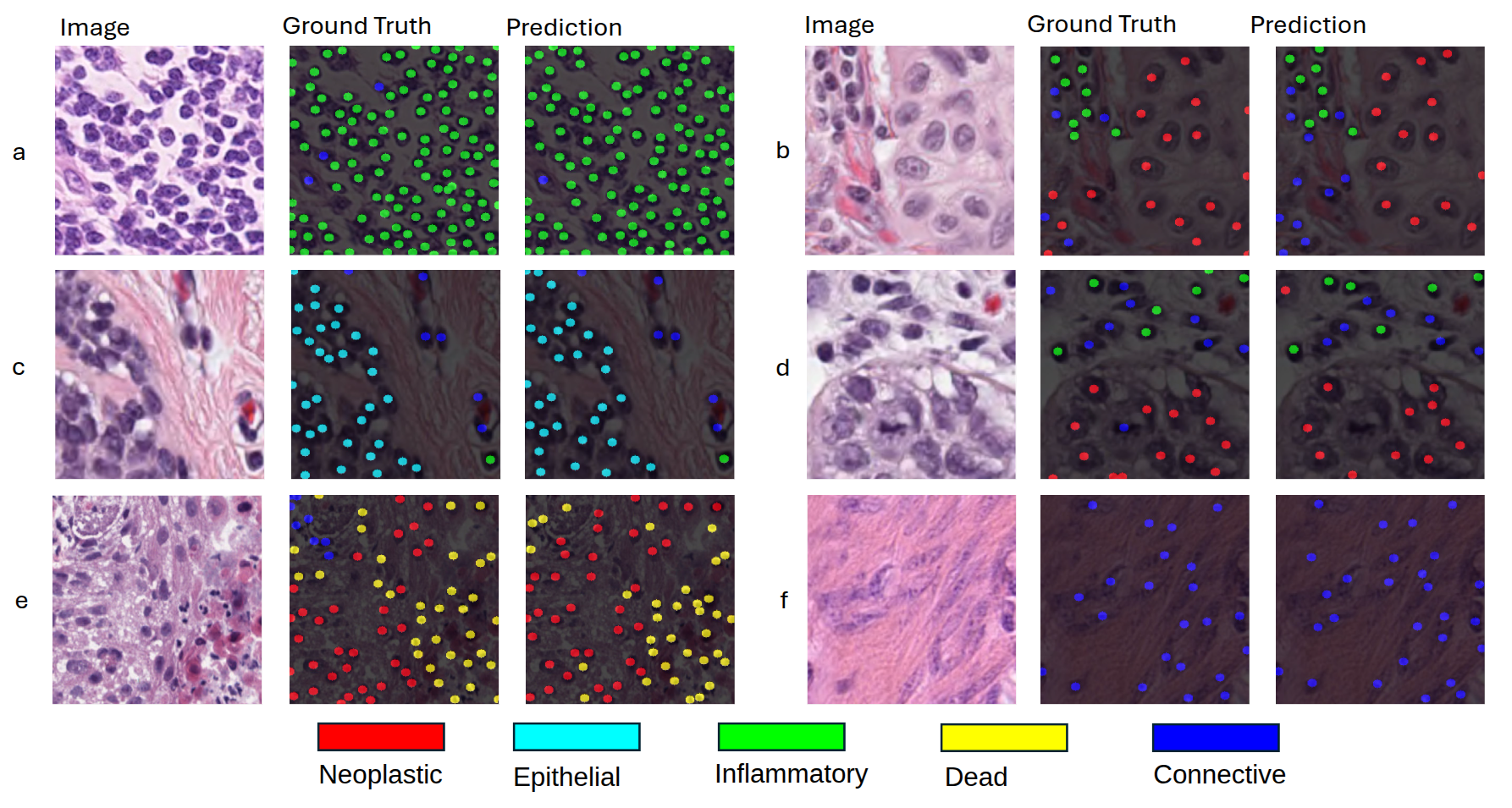}
    \caption{Examples from the PanNuke test set showing the input image, ground truth, and KongNet predictions.}
    \label{pannuke_pred_vis}
\end{figure}

Compared to other established methods such as STARDIST and HoVer-Net, KongNet shows a clear advantage in both overall and per-class performance. The strong results underscore the effectiveness of our architecture, which significantly outperforms transformer-based models such as CellViT that do not leverage massive pre-training, confirming the importance of task-specific architectural design. The qualitative results in Figure~\ref{pannuke_pred_vis} further illustrate KongNet's strong and balanced detection quality.

\begin{table}[t!]
\centering
\caption{Summary of detection results (F1) on the PanNuke Dataset for each cell type and overall detection. Neo: Neoplastic; Inf: Inflammatory; Epi: Epithelial; Con: Connective. Class average: the average of the F1 scores across five cell types.}
\label{pannuke_table}
\small
\begin{tabularx}{\textwidth}{
>{\raggedright\arraybackslash}p{3.8cm}
>{\centering\arraybackslash}p{0.8cm}
>{\centering\arraybackslash}p{0.8cm}
>{\centering\arraybackslash}p{0.8cm}
>{\centering\arraybackslash}p{0.8cm}
>{\centering\arraybackslash}p{0.8cm}
>{\centering\arraybackslash}p{0.8cm}
>{\centering\arraybackslash}X
}
\toprule
Method & Overall & Neo & Inf & Epi & Con & Dead & Class Average $\downarrow$ \\
\midrule
KongNet & \textbf{0.84} & 0.71 & \textbf{0.72} & 0.65 & \textbf{0.70} & \textbf{0.59} & \textbf{0.674} \\
CellNuc-DETR~\cite{cellnucdetr} & \textbf{0.84} & \textbf{0.72} & 0.60 & \textbf{0.74} & 0.56 & 0.45 & 0.618 \\
NuLite-H~\cite{NuLite} & 0.83 & 0.71 & 0.58 & 0.73 & 0.53 & 0.37 & 0.584 \\
CellViT-SAM-H~\cite{cellvit} & 0.83 & 0.71 & 0.58 & 0.73 & 0.53 & 0.36 & 0.582 \\
HoVer-NeXt~\cite{hovernext} & 0.76 & 0.61 & 0.66 & 0.55 & 0.56 & 0.53 & 0.582 \\
DualU-Net~\cite{dualunet} & 0.80 & 0.66 & 0.58 & 0.61 & 0.52 & 0.36 & 0.548 \\
CellViT no pre-train~\cite{cellvit} & 0.80 & 0.64 & 0.55 & 0.72 & 0.48 & 0.31 & 0.540 \\
STARDIST~\cite{stardist} & 0.82 & 0.69 & 0.57 & 0.70 & 0.51 & 0.10 & 0.514 \\
HoVer-Net~\cite{hovernet} & 0.80 & 0.62 & 0.54 & 0.56 & 0.49 & 0.31 & 0.504 \\
\bottomrule
\end{tabularx}
\end{table}

\subsection{PUMA Challenge Results}
We further tested KongNet's adaptability in the 2025 PUMA Challenge, which involved detecting and classifying nuclei in H\&E-stained images of advanced melanoma. The challenge featured two tracks of increasing complexity.

\subsubsection{PUMA Track 1: 3-Class Detection}
Track 1 focused on detecting three broad categories: tumour, lymphocytes, and other cells. In the preliminary leaderboard, KongNet ranked first across all cell types, achieving F1 scores of 0.8103 for tumour cells, 0.8070 for lymphocytes, and 0.5006 for other cells. The results are shown in Table~\ref{puma_table_track_1_preliminary}.

On the final leaderboard (Table~\ref{puma_table_track_1}), KongNet remained highly competitive. With a macro-average F1 of 0.6466, our model placed third overall in a very competitive race. KongNet achieved the highest F1 score (0.6746) for lymphocyte detection among all submitted methods. The primary reason for this outcome was that we only fine-tuned KongNet on the PUMA dataset within a limited time frame, using weights previously trained on the MONKEY model for lymphocyte and monocyte detection. Furthermore, KongNet significantly outperformed both the official HoVer-NeXt baseline and a fine-tuned CellViT++ model, demonstrating its robust performance against other SOTA approaches.

\begin{table}[t!]
\centering
\caption{Summary of detection results (F1) in the PUMA Challenge Track 1 \textbf{Final Leaderboard}. *Methods have not been publicly released at the time of writing. \textsuperscript{1}Multi-Stage Network~\cite{TeamLSM}. \textsuperscript{2}HoVer-NeXt model trained by challenge organisers~\cite{ref_puma_paper}. \textsuperscript{3}Team UME fine-tuned CellViT++ and performed a hyperparameter sweep~\cite{cellvit++puma}.}
\label{puma_table_track_1}
\begin{tabularx}{\columnwidth}{
>{\raggedright\arraybackslash}p{3.5cm}
>{\centering\arraybackslash}X
>{\centering\arraybackslash}X
>{\centering\arraybackslash}X
>{\centering\arraybackslash}X
}
\toprule
Method & Average $\downarrow$ & Tumour & Lymphocytes & Other \\
\midrule
Team rictoo* & \textbf{0.6585} & \textbf{0.8210} & 0.6729 & 0.4818 \\
Team LSM\textsuperscript{1} & 0.6501 & 0.7991 & 0.6537 & \textbf{0.4977} \\
KongNet & 0.6466 & 0.7948 & \textbf{0.6746} & 0.4704 \\
Baseline\textsuperscript{2} & 0.5979 & 0.7282 & 0.6611 & 0.4046 \\
Team UME\textsuperscript{3} & 0.5863 & 0.7614 & 0.6278 & 0.3697 \\
\bottomrule
\end{tabularx}
\end{table}

\subsubsection{PUMA Track 2}
Track 2 substantially increased the difficulty, requiring classification of ten distinct cell types. Again, KongNet ranked first overall in the preliminary phase. The results are shown in Table~\ref{puma_table_track_2_preliminary}. 

In the final test phase, KongNet secured a close second place overall with a macro-average F1 score of 0.2656, narrowly behind the winning team (0.2707), but much superior to other methods such as HoVer-NeXt and CellViT++ as shown in Table~\ref{puma_table_track_2}. The performance of our model was comparable to or exceeded the leading teams in most categories. KongNet again achieved the highest score for lymphocyte detection (0.6642) and also excelled at detecting melanophages and plasma cells compared to other methods. The performance gap to the winner (Team LSM) was primarily in rare or ambiguous classes like apoptotic cells, where their multi-stage pipeline, which incorporates tissue-level context, likely provided an advantage. This highlights a potential avenue for future improvement by integrating contextual information.

\begin{figure}[t!]
    \centering
    \includegraphics[width=\linewidth]{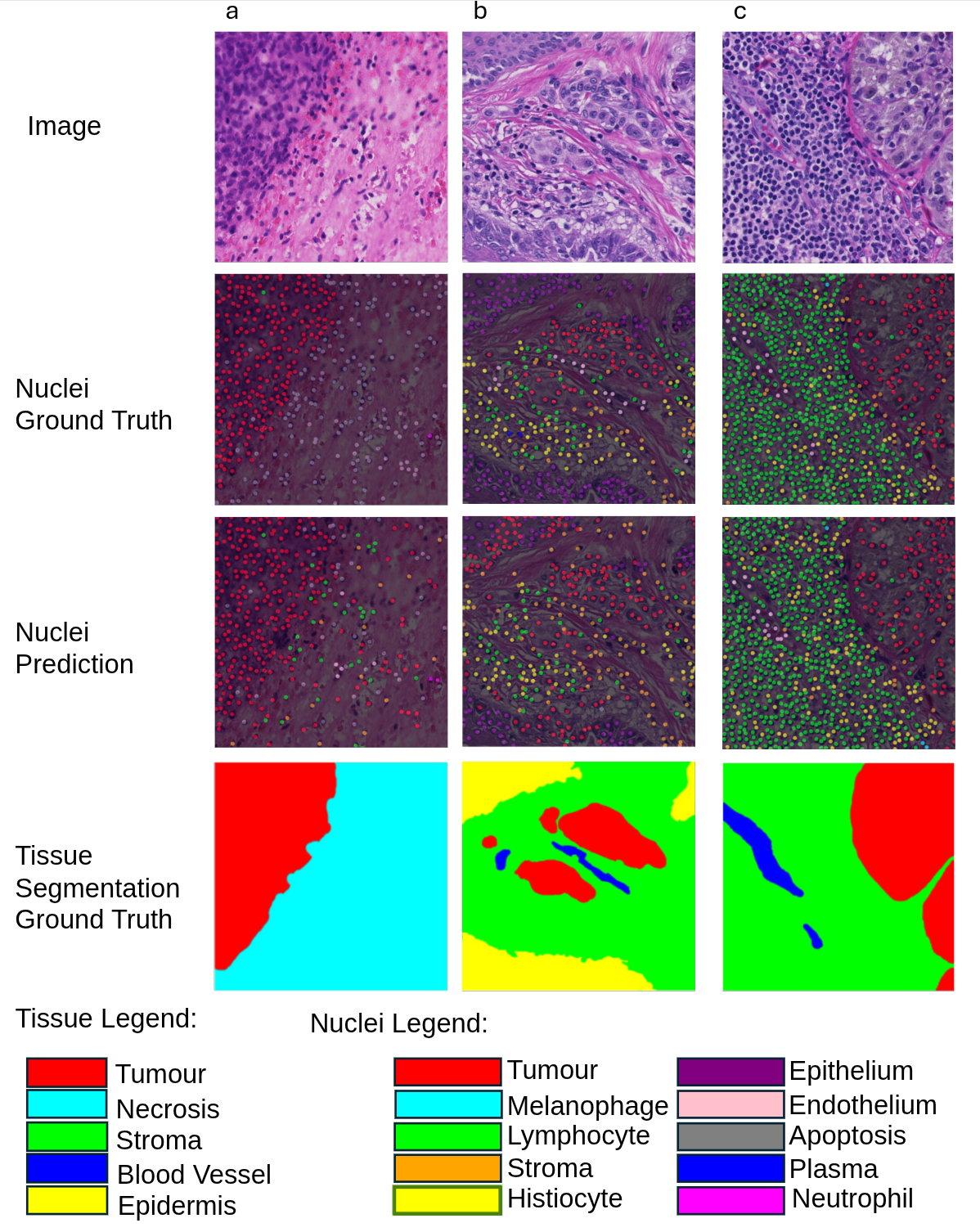}
    \caption{Visualization of detection quality across ten nuclei classes from our cross-validation results. a. KongNet confuses apoptotic cells with tumour cells and types of immune cells in the necrotic region. b. KongNet struggles to distinguish between tumour cells and epithelial cells in epidermis regions, suggesting that a larger context might be helpful for accurate cell classification. c. KongNet correctly detects and classifies most cells in the region.}
    \label{puma_pred_vis}
\end{figure}

A qualitative review of our model's failure cases (Figure~\ref{puma_pred_vis}) confirms these findings. The model occasionally confused apoptotic cells with lymphocytes and struggled to differentiate between tumour and epithelial cells, particularly in the absence of broader tissue context. This is likely because apoptotic cells often exhibit fragmented nuclei, which can resemble the small and rounded nuclei of lymphocytes. Another contributing factor may be the limited number of training examples containing necrotic regions. We also observed that distinguishing between tumour and epithelial cells can be challenging (Figure~\ref{puma_pred_vis}.b), as both share similar morphological features. These results indicate the potential value of incorporating tissue-type context into cell classification. For example, necrotic regions are more likely to contain apoptotic cells, while epidermal regions predominantly consist of epithelial cells rather than tumour cells.

\begin{table}[t!]
\centering
\caption{Summary of detection results (F1) in the PUMA Challenge Track 2 \textbf{Final Test Set}. *Methods have not been publicly released at the time of writing. \textsuperscript{1}Multi-Stage Network~\cite{TeamLSM}. \textsuperscript{2}HoVer-NeXt model trained by challenge organisers~\cite{ref_puma_paper}. \textsuperscript{3}Team UME fine-tuned CellViT++ and performed a hyperparameter sweep~\cite{cellvit++puma}.}
\label{puma_table_track_2}

\begin{subtable}{\textwidth}
\centering
\begin{tabularx}{\textwidth}{
>{\raggedright\arraybackslash}p{3.0cm}
*{6}{>{\centering\arraybackslash}X}
}
\toprule
Method & Average $\downarrow$ & Tum & Str & Apo & Epi & His \\
\midrule
Team LSM\textsuperscript{1} & \textbf{0.2707} & 0.7935 & \textbf{0.2978} & \textbf{0.1641} & \textbf{0.0817} & 0.2401 \\
KongNet & 0.2656 & 0.7952 & 0.2927 & 0.1170 & 0.0707 & 0.2154 \\
Team rictoo* & 0.2416 & \textbf{0.8190} & 0.2319 & 0.0394 & 0.0809 & \textbf{0.2413} \\
Team UME\textsuperscript{3} & 0.2335 & 0.7574 & 0.2500 & 0.1116 & 0.0632 & 0.2064 \\
Baseline\textsuperscript{2} & 0.2040 & 0.6795 & 0.1980 & 0.0262 & 0.0056 & 0.1885 \\
\bottomrule
\end{tabularx}
\caption*{Part 1: Average, Tum: Tumour, Str: Stroma, Apo: Apoptosis, Epi: Epithelium, His: Histiocyte.}
\end{subtable}

\vspace{0.5em}

\begin{subtable}{\textwidth}
\centering
\begin{tabularx}{\textwidth}{
>{\raggedright\arraybackslash}p{3.0cm}
*{5}{>{\centering\arraybackslash}X}
}
\toprule
Method & Lym & Neu & End & Mel & Pla \\
\midrule
Team LSM\textsuperscript{1} & 0.6299 & 0.0688 & 0.2053 & 0.1795 & 0.0465 \\
KongNet & \textbf{0.6642} & 0.0361 & 0.2123 & \textbf{0.1931} & \textbf{0.0595} \\
Team rictoo* & 0.6136 & \textbf{0.0068} & \textbf{0.2359} & 0.1474 & 0.0 \\
Team UME\textsuperscript{3} & 0.6107 & 0.0220 & 0.1421 & 0.1494 & 0.0218 \\
Baseline\textsuperscript{2} & 0.5847 & 0.0402 & 0.1553 & 0.1200 & 0.0422 \\
\bottomrule
\end{tabularx}
\caption*{Part 2: Lym: Lymphocyte, Neu: Neutrophil, End: Endothelium, Mel: Melanophage, Pla: Plasma.}
\end{subtable}

\end{table}

\subsection{MIDOG Challenge 2025 Results}
To validate the effectiveness and efficiency of our architectural principles on a completely different task, we employed our lightweight variant KongNet-Det, for mitosis detection during the MIDOG 2025 Challenge. This evaluation served as a rigorous benchmark of KongNet-Det in a highly competitive and challenging environment.

\subsubsection{Preliminary Leaderboard}
As shown in Table~\ref{midog_table_preliminary}, in the preliminary phase, KongNet-Det immediately established itself as a top contender, ranking first with an F1 score of 0.8147. Notably, our model achieved a good balance between precision (0.7995) and recall (0.8306) compared to other leading methods, which often skewed towards one or the other.

\subsubsection{MIDOG Final Leaderboard}
Table~\ref{midog_table_final} shows the performance on the final leaderboard. KongNet-Det once again secured first place with an F1 score of 0.7400. This is particularly significant given the difficulty of the official test set, which was designed to test the robustness of the models by including 12 different tumour types and highly challenging regions of interest (ROIs).

Achieving the top-rank in such a demanding benchmark underscores the power and efficiency of our lightweight architecture, which forms the core component of KongNet. It shows that the core design principles of KongNet are not limited to inflammatory cell detection and classification, but are versatile and can be extended to excel in other cell detection tasks such as mitosis detection.

\begin{table}[t!]
\centering
\caption{Summary of detection results (F1) in the MIDOG Challenge 2025 Track 1 \textbf{Final Leaderboard}. \textsuperscript{1}YOLOv12~\cite{midog_CBIO}. \textsuperscript{2}SDF-YOLO~\cite{midog_YTU}. \textsuperscript{3}YOLOv10~\cite{midog_cacfek}. \textsuperscript{4}RTMDet~\cite{midog_Gestalt}.}
\label{midog_table_final}
\begin{tabularx}{\columnwidth}{
>{\raggedright\arraybackslash}p{4.0cm}
>{\centering\arraybackslash}X
>{\centering\arraybackslash}X
>{\centering\arraybackslash}X
}
\toprule
Method & F1 $\downarrow$ & Precision & Recall \\
\midrule
KongNet-Det & \textbf{0.7400} & 0.7775 & 0.7059 \\
Team CBIO\textsuperscript{1} & 0.7216 & 0.7731 & 0.6765 \\
Team YTUVisionRG\textsuperscript{2} & 0.7085 & 0.6961 & \textbf{0.7215} \\
Team cacfek\textsuperscript{3} & 0.7063 & 0.7028 & 0.7098 \\
Team Gestalt\textsuperscript{4} & 0.7003 & \textbf{0.7973} & 0.6244 \\
\bottomrule
\end{tabularx}
\end{table}

\subsection{Ablation Studies}
We conducted a series of ablation studies on both the MONKEY and PanNuke datasets to validate two key design choices: the use of a multi-headed, specialised decoder architecture and the inclusion of SCSE attention modules.

\subsubsection{Impact of the Multi-Decoder Architecture}
First, we evaluated the performance of our main multi-decoder model (KongNet) against its simpler, single-decoder variants (KongNet-SH and KongNet-Det).

The results consistently demonstrate the advantage of using specialised decoders for complex classification tasks. On the MONKEY dataset, the full KongNet significantly outperformed both single-decoder models across all metrics (Table~\ref{decoder_ablation_monkey_table}). This trend was confirmed on the more complex PanNuke dataset (Table~\ref{decoder_ablation_pannuke_table}). While the single-head variant (KongNet-SH) achieved the same high overall detection F1 score (0.84), the multi-headed KongNet achieved the highest class-average F1 score (0.674). This indicates that although a single decoder is highly effective for general nuclei detection, the specialised decoders are beneficial for the complex task of accurately distinguishing between multiple cell types.

\subsubsection{Impact of SCSE Attention Modules}
Next, we assessed the benefit of integrating SCSE attention modules within the decoders. The findings confirm their positive impact on performance. For our main KongNet architecture, adding SCSE modules consistently improved performance on the MONKEY dataset, for example, raising the FROC score of monocytes from 0.2170 to 0.2407 (Table~\ref{decoder_ablation_monkey_table}). On the PanNuke dataset, for example, adding SCSE modules to KongNet increased the class-average F1 score from 0.656 to 0.674 (Table~\ref{decoder_ablation_pannuke_table}). While the improvements were less pronounced for the single-decoder variants, the consistent gains observed for our primary KongNet architecture confirmed their importance. Based on these findings, SCSE modules were integrated into our final model configuration, contributing to its strong and robust performance.

\begin{table}[t!]
\centering
\caption{Summary of detection results (FROC) from our internal 5-fold cross-validation on the MONKEY training dataset. Ablation study on the effects of the multi-decoder architecture and SCSE modules.}
\label{decoder_ablation_monkey_table}
\begin{tabularx}{\columnwidth}{
>{\raggedright\arraybackslash}p{2.5cm}
>{\centering\arraybackslash}X
>{\centering\arraybackslash}X
>{\centering\arraybackslash}X
}
\toprule
Method & Inflammatory Cells & Lymphocytes & Monocytes \\
\midrule
KongNet-Det No SCSE & $0.3053 \pm 0.0410$ & $0.3414 \pm 0.0451$ & $0.1768 \pm 0.0199$ \\
KongNet-Det & $0.3114 \pm 0.0326$ & $0.3543 \pm 0.0178$ & $0.2140 \pm 0.0352$ \\
\midrule
KongNet-SH No SCSE & $0.3194 \pm 0.0242$ & \bm{$0.3951 \pm 0.0292$} & $0.2222 \pm 0.0046$ \\
KongNet-SH & $0.3263 \pm 0.0406$ & $0.3902 \pm 0.0376$ & $0.2208 \pm 0.0171$ \\
\midrule
KongNet No SCSE & $0.3408 \pm 0.0306$ & $0.3896 \pm 0.0117$ & $0.2170 \pm 0.0161$ \\
KongNet & \bm{$0.3537 \pm 0.0306$} & $0.3950 \pm 0.0378$ & \bm{$0.2407 \pm 0.0084$} \\
\bottomrule
\end{tabularx}
\end{table}

\begin{table}[t!]
\centering
\caption{Summary of detection results (F1) on the PanNuke Dataset for each cell type and overall detection. Ablation study on the effects of the multi-decoder architecture and SCSE modules. Neo: Neoplastic; Inf: Inflammatory; Epi: Epithelial; Con: Connective. Class average: the average F1 scores across five cell types.}
\label{decoder_ablation_pannuke_table}
\begin{tabularx}{\textwidth}{
>{\raggedright\arraybackslash}p{3.8cm}
>{\centering\arraybackslash}p{1.0cm}
>{\centering\arraybackslash}p{0.8cm}
>{\centering\arraybackslash}p{0.8cm}
>{\centering\arraybackslash}p{0.8cm}
>{\centering\arraybackslash}p{0.8cm}
>{\centering\arraybackslash}p{0.8cm}
>{\centering\arraybackslash}X
}
\toprule
Method & Overall & Neo & Inf & Epi & Con & Dead & Class Average\textsuperscript{*} \\
\midrule
KongNet-Det No SCSE & 0.83 & 0.69 & 0.71 & 0.63 & 0.67 & 0.58 & 0.656 \\
KongNet-Det & 0.83 & 0.69 & 0.72 & 0.61 & 0.66 & 0.58 & 0.652 \\
\midrule
KongNet-SH No SCSE & \textbf{0.84} & \textbf{0.71} & \textbf{0.73} & 0.63 & \textbf{0.70} & \textbf{0.60} & 0.674 \\
KongNet-SH & \textbf{0.84} & \textbf{0.71} & 0.72 & 0.63 & \textbf{0.70} & \textbf{0.60} & 0.672 \\
\midrule
KongNet No SCSE & \textbf{0.84} & 0.70 & 0.71 & 0.60 & 0.69 & 0.58 & 0.656 \\
KongNet & \textbf{0.84} & \textbf{0.71} & 0.72 & \textbf{0.65} & \textbf{0.70} & 0.59 & \textbf{0.674} \\
\bottomrule
\end{tabularx}
\end{table}

\section{Inference Speed and Efficiency}
\subsection{Whole-Slide Image Inference Runtime}
To assess practical deployment efficiency, we benchmarked the end-to-end inference time for a full whole-slide image (WSI). The benchmark was performed at $40\times$ magnification using a TCGA slide (ID:\texttt{TCGA-AN-A0XP-01Z-00-DX1}), which contains approximately $274~\text{mm}^2$ of foreground tissue. We compared KongNet against HoVer-NeXt Tiny and CellViT-SAM-H on two hardware tiers: a standard workstation and a high-performance computing (HPC) node, reflecting different resource environments in pathology labs.

\textbf{Hardware specifications}:
\begin{itemize}
\item \textbf{Workstation}: NVIDIA RTX 3060 (12 GB), Intel Core i5-13500 (14 cores), 32 GB RAM
\item \textbf{HPC Node}: NVIDIA V100 (32 GB), Intel Xeon Platinum 8168 (32 cores), 80 GB RAM
\end{itemize}

The CellViT inference pipeline automatically selects batch size and the number of worker threads based on available hardware. HoVer-NeXt Tiny was configured with \texttt{pp\_tiling=16}, \texttt{inf\_workers=10}, and \texttt{pp\_workers=10} on the workstation, and with \texttt{inf\_workers=32} and \texttt{pp\_workers=32} on the HPC node.

On the workstation (Table~\ref{inference_time_40x_workstation}), KongNet demonstrated excellent efficiency. Without test-time augmentation (TTA), it processed the WSI in just 8 minutes. Using 4× TTA, which offers the best balance of accuracy and speed, the runtime was 26 minutes, $16\times$ TTA increases the runtime to 94 minutes. HoVer-NeXt Tiny ($16\times$ TTA) completes inference in 31 minutes, we used $16\times$ TTA as this setting achieves the reported detection performance. The 699-million-parameter CellViT-SAM-H failed to run on the workstation due to memory errors, highlighting the hardware barrier large transformer models present for deployment in typical, resource-constrained settings.

On the HPC node (Table~\ref{inference_time_40x_hpc}), KongNet requires only 6 minutes without TTA, 14 minutes with $4\times$ TTA, and 43 minutes with $16\times$ TTA. CellViT-SAM-H was successfully evaluated on the HPC node, requiring 18 minutes total, $3\times$ longer than KongNet with no TTA. 

In summary, although in our main results section we report the performance of KongNet with $16\times$ TTA, running KongNet with $4\times$ TTA could offer the best trade-off between detection quality and computational efficiency, which is a more practical choice for real-world applications. We found that $16\times$ TTA yields minimal accuracy gains over $4\times$ TTA on PanNuke dataset (Table~\ref{tta_ablation_pannuke_table})

\subsection{Isolated Model Forward Pass Latency}
To measure raw model speed independent of the processing pipeline, we benchmarked the average forward pass time on the workstation hardware. As expected, the much smaller HoVer-NeXt Tiny was the fastest, with a latency of just 0.0133 seconds. KongNet's forward pass (0.0378 s) was approximately four times faster than CellViT-SAM-H (0.1510 s) and slightly faster than the smaller CellViT 256 (0.0384 s) despite having 3.8 times more parameters.
This suggests that the architectural design of KongNet, being a convolutional neural network (CNN), benefits from hardware-level optimisations compared to certain operations in Vision Transformers (ViTs) such as self-attention and LayerNorm, which can be computationally intensive. The results are summarised in Table~\ref{forward_pass}.

\section{Discussion}

\subsection{CNNs vs ViT}

Our results highlight the distinct, complementary strengths of Convolutional Neural Networks (CNNs) and Vision Transformers (ViTs) for nuclei detection and classification. ViTs excel at modelling long-range spatial dependencies through self-attention, making them adept at integrating tissue context. In contrast, CNNs remain unparalleled at learning the fine-grained local morphological features that are often the primary discriminative information in histopathology. To further investigate this point within our own framework, we compared CNN- and transformer-based encoder backbones for KongNet on the MONKEY dataset. As shown in Table~\ref{ablation_vit_encoder_monkey} of the Supplementary Material, the CNN-based EfficientNetV2-L encoder achieved the strongest overall performance.

This dichotomy is evident in the PanNuke results (Table~\ref{pannuke_table}). Our CNN-based KongNet excelled at classifying cell types defined by their distinct nuclear morphology, such as inflammatory, connective, and dead cells. ViT-based models such as CellViT and CellNuc-DETR showed an advantage in classifying neoplastic and epithelial nuclei categories, where the surrounding tissue architecture provides a crucial context to which ViTs are better suited. The MONKEY Challenge results (Table~\ref{monkey_table_final}) further support this: in a task where subtle morphological differences between lymphocytes and monocytes are key and broader context is less informative, CNN-based models, including ours, outperformed ViT-based approaches.

These findings suggest a task-dependent synergy. CNN-based architectures such as KongNet appear optimal when classification is based on the morphology of the nucleus itself, while ViTs may hold an advantage when the task relies on interpreting broader spatial context. This points toward a promising future direction: hybrid architectures that skilfully combine the local feature extraction power of CNNs with the global context modelling of ViTs.

\subsection{Impact of the Multi-headed Architecture}

To better understand why the relative performance of the single-head and multi-head variants differs across datasets, we investigated whether this behaviour is driven by the intrinsic characteristics of the datasets. Overall, our results suggest that the dominant factor is the dataset characteristics, in particular the degree of morphological separability between cell classes.

On the MONKEY dataset, the multi-head architecture provides a clear advantage over the single-head variant across all three cell categories. KongNet-SH achieved FROC scores of 0.3263, 0.3902, and 0.2208 for inflammatory cells, lymphocytes, and monocytes, respectively, whereas the multi-head KongNet improved these to 0.3537, 0.3950, and 0.2407 (Table~\ref{monkey_table_local_validation}). In contrast, on the PanNuke dataset, KongNet and KongNet-SH achieved very similar overall performance, with class-average F1 scores of 0.674 and 0.672, respectively (Table~\ref{decoder_ablation_pannuke_table}). These findings indicate that the benefit of class-specific decoders is dependent on the difficulty of the underlaying cell classification problem.

We hypothesised that the tasks requiring separation of highly overlapping cell morphologies benefit more from a multi-headed architecture, whereas tasks in which the target cell classes are more readily separable gain smaller benefit from decoder specialisation. In particular, the MONKEY dataset requires separating lymphocytes and monocytes, which exhibit highly similar morphology, while the PanNuke dataset contains several classes that are more readily distinguishable. To test this hypothesis, we analysed nuclei distributions in feature space for both datasets. For each dataset, we first extracted nuclei instances from the available segmentation masks (Examples are shown in Figure~\ref{monkey_pannuke_nuclei}). For MONKEY, we used NuClick generated masks employed during KongNet training, for PanNuke we used the provided instance annotations. Then, for each nucleus, we computed a set of hand-crafted shape descriptors (Supplementary Material~\ref{hand_crafted_features}) and extracted deep features using an ImageNet pretrained ResNet18 model. We then treated each cell type as a cluster in feature space and quantified cluster separability using the Davies--Bouldin (DB) index~\cite{DB_Index}, where lower values indicate more compact and better separated clusters.

In the MONKEY dataset, the two target classes, lymphocytes and monocytes, showed relatively poor separability, with DB indices of 6.74 in the shape-feature space and 7.57 in the ResNet18 feature space (Table~\ref{monkey_DB_index}). By contrast, in PanNuke, most pairwise class comparisons yielded substantially lower DB values, indicating better class separation overall. The only exception was the pair neoplastic versus epithelial cells, which exhibited the highest overlap in PanNuke, with DB indices of 11.61 and 8.32 for shape and ResNet18 features respectively (Table~\ref{pannuke_DB_index}). This is consistent with the observation that the multi-head design provides only a modest overall gain on PanNuke, with the most benefit arising in the detection of neoplastic and epithelial cells, thus supporting our hypothesis.

Figure~\ref{tsne_plot} provides a qualitative visualisation of clusters. In the MONKEY dataset, the lymphocyte and monocyte clusters show substantial overlap, consistent with their relatively high DB indices. In the PanNuke dataset, the neoplastic and epithelial classes also appear less well separated than other pairs such as dead and neoplastic classes, which visually reflects the corresponding DB values. Although \textit{t}-SNE is only a qualitative projection, it provides an intuitive illustration of the same trend captured quantitatively by the DB index.

To further examine the role of loss weighting, we evaluated automatic class-wise loss weighting on the PanNuke dataset using the same strategy employed for MONKEY. As shown in Table~\ref{ablation_pannuke_auto_lambda}, automatic $\lambda$ weighting reduced the class-average F1 score of KongNet from 0.674 to 0.654, indicating that it is inferior to the fixed equal-weighting scheme. We therefore used equal $\lambda$ values for PanNuke in all main comparisons, as this provided the strongest performance for the model. Consequently, the comparison between KongNet-SH and KongNet-MH on PanNuke was already performed under the optimal loss-weighting setting we identified, indicating that the small difference between the two variants is unlikely to be explained by the choice of $\lambda$ values.

Together, these findings suggest that the effectiveness of the multi-headed design depends on the task and the structure of the dataset. When the target classes are morphologically similar and occupy strongly overlapping regions of feature space, as in the MONKEY dataset, independent decoder heads appear to provide an improved class separation. When the classes are already relatively well separated, as for most the PanNuke dataset categories, the advantage of the multi-headed design is smaller.

\subsection{Tissue-Guided Cell Classification}
Previous work, such as OCELOT~\cite{ocelot}, has shown that incorporating tissue-level context can enhance cell classification accuracy, particularly in complex tumour microenvironment. However, the OCELOT dataset was limited to only two types of classifications, tumour vs non-tumour cells and tissue.

The PUMA Challenge results underscore the important role of tissue context in resolving cellular ambiguity. The winning team (Team LSM~\cite{TeamLSM}) employed a multi-stage pipeline that used tissue segmentation maps to guide and refine nuclei classification, allowing them to outperform our model. Their advantage was most pronounced for apoptotic cells, a morphologically ambiguous category. As shown in our qualitative analysis and supported by their results (Table~\ref{puma_table_track_2}), apoptotic cells frequently occur within necrotic tissue regions. A model aware of this necrotic context is more likely to correctly classify ambiguous nuclear fragments as apoptotic, whereas a context-agnostic model like ours may confuse them with immune cells such as lymphocytes.

This strongly suggests that the next major advance in cell classification, particularly for ambiguous or transitional cell states, lies in joint tissue-cell modelling. While KongNet in its current form does not leverage this information, a clear and promising path for future work is to integrate tissue context directly into the architecture. This could allow the model to make more informed, pathologist-like decisions in challenging histopathological settings.

\section{Conclusion}
In this work, we introduced KongNet, a multi-headed deep learning architecture designed for accurate and generalisable cell detection and classification in histopathology. The model's core contribution is its use of parallel, cell-type-specialised decoders, enhanced by SCSE attention and a multi-task learning framework that jointly predicts centroids, segmentation masks, and contours. This design proved highly effective at learning discriminative morphological features while minimising inter-class interference.

We demonstrated the power and robustness of this approach through a comprehensive evaluation across multiple Grand Challenges and public benchmarks. KongNet and its lightweight variant, KongNet-Det, achieved first-place victories in the MONKEY Challenge for inflammatory cell detection and the 2025 MIDOG Challenge for mitosis detection, respectively. Moreover, KongNet also secured top-three rankings in the highly complex PUMA Challenge and established new state-of-the-art performance on the PanNuke and CoNIC datasets, proven its effectiveness across diverse tissues, cell types, and staining protocols.

In summary, KongNet provides a robust, adaptable, and highly competitive architectural foundation for a wide range of nuclei analysis tasks. Its consistent state-of-the-art performance establishes it as a powerful and practical tool for the computational pathology community.

Future work will focus on leveraging this tool for large-scale analysis of the tumour microenvironment (TME). We plan to create a comprehensive pipeline by combining KongNet with our first-place-winning tissue segmentation model~\cite{puma_tissue_model} from the PUMA Challenge. Furthermore, inspired by our findings in the Discussion, we will explore next-generation architectures incorporating contextual information that explicitly integrates tissue context to further enhance classification accuracy in the most challenging histopathological settings.

\section*{Code Availability}
\begin{itemize}
    \item Full inference code can be accessed on GitHub at this link \url{https://github.com/Jiaqi-Lv/KongNet_Inference_Main}.
    \item The model is also accessible as a Jupyter Notebook in TIAToolBox~\cite{tiatoolbox}, which can be accessed at this link \url{https://tia-toolbox.readthedocs.io/en/latest/_notebooks/jnb/12-nucleus-detection.html}.
\end{itemize}

\section*{Acknowledgement} JL is supported by the UK Engineering and Physical Sciences Research Council (EPSRC). SEAR reports financial support by the MRC (MR/X011585/1) and the BigPicture project,  which has received funding from the Innovative Medicines Initiative 2 Joint Undertaking under grant agreement No 945358. We thank the organisers and contributors of the MONKEY Challenge, the CoNIC Challenge, the PUMA Challenge, and the 2025 MIDOG challenge, for providing the valuable datasets and evaluation platforms.

\clearpage

\bibliographystyle{elsarticle-harv} 
\bibliography{main.bib}

\clearpage
\appendix
\beginsupplement
Supplementary Material

\section{Supplementary Information}
\subsection{Summary of MONKEY Challenge Submission}\label{monkey_submission_desc}
We used MONKEY Dataset as the primary dataset to iteratively develop our model based on internal cross-validation. We configure KongNet with three decoders, one for overall inflammatory cell detection, one for lymphocytes detection, and one for monocytes detection.

\subsection{Summary of PUMA Challenge Submission}\label{puma_submission_desc}
We evaluated the adaptability of our model through participation in the PUMA Challenge, which comprised two tracks focusing on nuclei detection and classification in advanced melanoma:  
\textbf{Track 1:} Detection of three types of cells: tumour cells, tumour-infiltrating lymphocytes (TILs; including lymphocytes and plasma cells), and other cells.  
\textbf{Track 2:} Detection of ten types of cells: tumour cells, lymphocytes, plasma cells, histiocytes, melanophages, neutrophils, stromal cells, epithelium cells, endothelium cells, and apoptotic cells.

Adapting our model to the PUMA Challenge required only a single modification: adjusting the number of decoders. We used three decoders for Track~1 and ten decoders for Track~2. Due to time constraints, we did not perform hyperparameter tuning.

\subsection{Summary of 2025 MIDOG Challenge Submission}\label{midog_submission_desc}
Given the condensed timeline for our participation in the MIDOG 2025 Challenge, we strategically chose to deploy our lightweight and efficient KongNet-Det model for the mitosis detection task. This provided an ideal opportunity to validate our streamlined architecture on a large-scale, external benchmark. For this challenge, KongNet-Det was configured with a single output channel to predict the centroids of mitotic figures, the model was initialised with PanNuke pre-trained weights.

\subsection{Summary of PanNuke and CoNIC Model Configurations}
In the PanNuke dataset, five types of cells are of interest: neoplastic, inflammatory, connective, dead, and non-neoplastic epithelial cells. For this task, we configured KongNet with six decoders, one for each of the five cell types, plus an additional decoder for overall cell detection. 

In the CoNIC dataset, the six types of cells of interest are: neutrophils, lymphocytes, eosinophils, connective tissue cells, epithelial cells, and plasma cells. Accordingly, we used six separate decoders, one for each cell type. 

\subsection{Training Hyperparameters}
We summarise the key training and post-processing hyperparameters for each dataset below. Unless otherwise stated, we use the AdamW optimiser with an initial learning rate of $4\times10^{-4}$ and apply learning rate scheduling. Post-processing follows a standard pipeline involving peak detection and non-maximum suppression (NMS), with dataset-specific parameters. 

We employ data augmentation techniques during training to ensure that the model is robust against stain variations and prevent overfitting. The following methods from Albumentations~\cite{ablumentations} are used: RGB shift, HSV shift, Gaussian Blur, Sharpen, Image Compression, Random Brightness Contrast, Shift Scale Rotate. We apply further augmentations by employing Strong Augmentation~\cite{strong_augment} with their default settings.

\subsubsection{MONKEY Challenge Dataset}

\begin{itemize}
    \item \textbf{Batch size:} 48
    \item \textbf{Learning rate schedule:} CosineAnnealingWarmRestarts
    \item \textbf{Weight decay:} 0.01
    \item \textbf{Centroid dilation:} Circular disks with diameter 11 pixels
    \item \textbf{Post-processing:}
    \begin{itemize}
        \item Output Probability = $0.6\times$ Centroid probability + $0.4\times$ segmentation probability
        \item Peak Local Max threshold: 0.5, minimum distance: 11 px
        \item NMS overlap threshold: 0.5, box size: 11 px
    \end{itemize}
\end{itemize}

\subsubsection{PUMA Challenge Dataset}

\begin{itemize}
    \item \textbf{Batch size:} 48 (Track 1), 24 (Track 2)
    \item \textbf{Learning rate schedule:} CosineAnnealingLR
    \item \textbf{Weight decay:} 0.01
    \item \textbf{Centroid dilation:} Circular disks with diameter 11 pixels
    \item \textbf{Post-processing:}
    \begin{itemize}
        \item Output Probability = $0.6\times$ Centroid probability + $0.4\times$ segmentation probability
        \item Peak Local Max threshold: 0.5, minimum distance: 11 px
        \item NMS overlap threshold: 0.5, box size: 11 px
    \end{itemize}
\end{itemize}

\subsubsection{PanNuke Dataset}

\begin{itemize}
    \item \textbf{Batch size:} 32
    \item \textbf{Learning rate schedule:} CosineAnnealingLR
    \item \textbf{Weight decay:} 0.05
    \item \textbf{Centroid dilation:} Circular disks with diameter 9 pixels
    \item \textbf{Post-processing:}
    \begin{itemize}
        \item Output Probability = Centroid probability.
        \item Peak Local Max threshold: 0.5, minimum distance: 9 px
        \item NMS overlap threshold: 0.5, box size: 9 px
    \end{itemize}
\end{itemize}

\subsubsection{CoNIC Dataset}

\begin{itemize}
    \item \textbf{Batch size:} 32
    \item \textbf{Learning rate schedule:} CosineAnnealingLR
    \item \textbf{Weight decay:} 0.05
    \item \textbf{Centroid dilation:} Circular disks with diameter 3 pixels
    \item \textbf{Post-processing:}
    \begin{itemize}
        \item Output Probability = Centroid probability.
        \item Peak Local Max threshold: 0.5, minimum distance: 3 px
        \item NMS overlap threshold: 0.5, box size: 3 px
    \end{itemize}
\end{itemize}

\subsubsection{MIDOG Challenge Dataset}
\begin{itemize}
    \item \textbf{Batch size:} 40
    \item \textbf{Learning rate schedule:} CosineAnnealingLR
    \item \textbf{Weight decay:} 0.01
    \item \textbf{Centroid dilation:} Circular disks with diameter 21 pixels
    \item \textbf{Post-processing:}
    \begin{itemize}
        \item Output Probability = Centroid probability.
        \item Peak Local Max threshold: 0.99, minimum distance: 21 px
        \item NMS overlap threshold: 0.5, box size: 21 px
    \end{itemize}
\end{itemize}

\subsubsection{Hand-crafted Morphological Features for Nuclei Analysis}
\label{hand_crafted_features}

For each segmented nucleus instance, we extracted a set of hand-crafted morphological features. These were used to analyse class separability in feature space.

\begin{itemize}
    \item \textbf{Area}: number of foreground pixels in the nucleus mask.
    \item \textbf{Perimeter}: boundary length of the nucleus mask.
    \item \textbf{Circularity} \((4\pi A / P^2)\): higher values indicate a more circle-like shape.
    \item \textbf{Compactness} \((P^2 / A)\): lower values indicate a more compact object.
    \item \textbf{Eccentricity}: measures how elongated the nucleus is.
    \item \textbf{Major axis length}: length of the major axis of the ellipse fitted from second-order moments.
    \item \textbf{Minor axis length}: length of the minor axis of the ellipse fitted from second-order moments.
    \item \textbf{Axis ratio} (minor/major): ratio between minor and major axis lengths.
    \item \textbf{Equivalent diameter}: diameter of a circle with the same area as the nucleus.
    \item \textbf{Convex area}: area of the convex hull of the nucleus.
    \item \textbf{Solidity} (area/convex area): measures how completely the nucleus fills its convex hull.
    \item \textbf{Convex hull perimeter}: perimeter of the convex hull of the nucleus.
    \item \textbf{Convexity} (convex hull perimeter / perimeter): compares the convex hull boundary with the original nucleus boundary.
    \item \textbf{Extent} (area/bounding-box area): proportion of the bounding box occupied by the nucleus.
    \item \textbf{Filled area}: area after internal holes are filled.
    \item \textbf{Euler number}: topological descriptor reflecting connected components and internal holes.
    \item \textbf{Radial mean}: mean distance from the nucleus centroid to boundary pixels.
    \item \textbf{Radial standard deviation}: standard deviation of distances from the centroid to the boundary.
    \item \textbf{Radial maximum}: maximum centroid-to-boundary distance.
    \item \textbf{Radial coefficient of variation}: ratio of radial standard deviation to radial mean.
    \item \textbf{Maximum Feret diameter}: maximum caliper diameter of the nucleus.
    \item \textbf{Hu moment 1 -- 7}
    \item \textbf{Log Hu moment 1 -- 7}: signed log-transformed version of Hu moment 1 -- 7.
    \item \textbf{Foreground fraction}: ratio of nucleus area to total cropped image area.
\end{itemize}

The Hu moments are invariant to translation, scale, and rotation. Together, these features capture complementary aspects of nucleus size, elongation, compactness, convexity, topology, boundary regularity, and global shape.

\clearpage

\begin{table}[ht!]
\centering
\caption{Summary of detection results (FROC) from our \textbf{internal 5-fold cross-validation} on the \textbf{MONKEY} training dataset.}
\label{monkey_table_local_validation}
\begin{tabularx}{\columnwidth}{
>{\raggedright\arraybackslash}p{2.9cm}
>{\centering\arraybackslash}X
>{\centering\arraybackslash}X
>{\centering\arraybackslash}X
}
\toprule
Method & Inflammatory Cells & Lymphocytes & Monocytes \\
\midrule
KongNet-Det & $0.3114 \pm 0.0326$ & $0.3543 \pm 0.0178$ & $0.2140 \pm 0.0352$ \\
KongNet-SH & $0.3263 \pm 0.0406$ & $0.3902 \pm 0.0376$ & $0.2208 \pm 0.0171$ \\
KongNet & $0.3537 \pm 0.0306$ & $0.3950 \pm 0.0378$ & \bm{$0.2407 \pm 0.0084$} \\
KongNet (wide) & \bm{$0.3641 \pm 0.0302$} & \bm{$0.4031 \pm 0.0315$} & $0.2359 \pm 0.0094$ \\
\bottomrule
\end{tabularx}
\end{table}

\begin{table}
\centering
\caption{Summary of detection results (FROC) on the MONKEY Challenge \textbf{Preliminary Leaderboard}. $^*$For other teams, we report the submission with the highest FROC score from the preliminary leaderboard; we cannot guarantee that the same algorithm was submitted to the final leaderboard by the other teams. $^1$Faster R-CNN trained by organisers of the MONKEY Challenge~\cite{ref_monkey_challenge}.}
\label{monkey_table_preliminary}
\begin{tabularx}{\columnwidth}{
>{\raggedright\arraybackslash}p{5cm}
>{\centering\arraybackslash}X
>{\centering\arraybackslash}X
>{\centering\arraybackslash}X
}
\toprule
Method & Inflammatory Cells $\downarrow$ & Lymphocytes & Monocytes \\
\midrule
Team InstanSeg$^*$ & \textbf{0.4184} & \textbf{0.4394} & \textbf{0.1998} \\
KongNet (Wide) & 0.3972 & 0.3996 & 0.1743 \\
Team AIRA Matrix$^*$ & 0.3633 & 0.3844 & 0.1316 \\
Team ST Medical$^*$ & 0.3274 & 0.3618 & 0.1104 \\
Team BioTotem$^*$ & 0.3103 & 0.3327 & 0.0726 \\
Team ouradiology$^*$ & 0.2938 & 0.3096 & 0.0774 \\
Team Zip Lab UNIMORE$^*$ & 0.2470 & 0.3049 & 0.0699 \\
Baseline$^1$ & 0.2186 & 0.2240 & 0.0749 \\
\bottomrule
\end{tabularx}
\end{table}

\begin{table}
\centering
\caption{Ablation study on the MONKEY dataset comparing CNN- and transformer-based encoder backbones for KongNet. Results are reported as mean FROC for inflammatory cells, lymphocytes, and monocytes across our internal 5-fold cross-validation on the MONKEY training set.}
\label{ablation_vit_encoder_monkey}
\begin{tabularx}{\linewidth}{
>{\raggedright\arraybackslash}p{2.9cm}
>{\centering\arraybackslash}X
>{\centering\arraybackslash}X
>{\centering\arraybackslash}X
}
\toprule
Method & Inflammatory Cells & Lymphocytes & Monocytes \\
\midrule
EfficientNetV2-L & \bm{$0.3537 \pm 0.0306$} & $0.3950 \pm 0.0378$ & \bm{$0.2407 \pm 0.0084$} \\
SwinV2-B & $0.3324 \pm 0.0027$ & $0.3727 \pm 0.0283$ & $0.1924 \pm 0.0144$ \\
ConvNeXtV2-L & $0.3418 \pm 0.0034$ & \bm{$0.3959 \pm 0.0333$} & $0.2198 \pm 0.0115$ \\
ConvNeXtV2-B & $0.3461 \pm 0.0332$ & $0.3859 \pm 0.0272$ & $0.2176 \pm 0.0120$ \\
\bottomrule
\end{tabularx}
\end{table}

\begin{table}
\centering
\caption{Ablation study on the MONKEY dataset evaluating different weightings of the auxiliary nuclei segmentation and contour segmentation losses in KongNet. Results are reported as mean FROC scores for inflammatory cells, lymphocytes, and monocytes across our internal 5-fold cross-validation. The class-specific loss for each cell class $k$ is defined as $L_{\mathrm{Class}_k} = L_{\mathrm{Centroid}_k} + \lambda_1 L_{\mathrm{Seg}_k} + \lambda_2 L_{\mathrm{Contour}_k}$, where $\lambda_1$ and $\lambda_2$ denote the weights of the segmentation and contour losses, respectively. The final KongNet model uses $\lambda_1 = 1.0$ and $\lambda_2 = 0.5$; setting $\lambda_1 = \lambda_2 = 0.0$ gives the detection-only KongNet-Det model.}
\label{ablation_monkey_loss_weighting}
\begin{tabularx}{\linewidth}{
>{\centering\arraybackslash}p{0.9cm}
>{\centering\arraybackslash}p{0.9cm}
>{\centering\arraybackslash}X
>{\centering\arraybackslash}X
>{\centering\arraybackslash}X
}
\toprule
$\lambda_1$ & $\lambda_2$ & Inflammatory Cells & Lymphocytes & Monocytes \\
\midrule
1.0 & 0.5 & \bm{$0.3537 \pm 0.0306$} & \bm{$0.3950 \pm 0.0378$} & \bm{$0.2407 \pm 0.0084$} \\
0.5 & 0.5 & $0.3428 \pm 0.0244$ & $0.3857 \pm 0.0254$ & $0.2166 \pm 0.0073$ \\
1.0 & 0.0 & $0.3473 \pm 0.0293$ & $0.3939 \pm 0.0290$ & $0.2165 \pm 0.0120$ \\
0.0 & 1.0 & $0.2777 \pm 0.0355$ & $0.3540 \pm 0.0257$ & $0.2055 \pm 0.0257$ \\
0.0 & 0.0 & $0.3114 \pm 0.0326$ & $0.3543 \pm 0.0178$ & $0.2140 \pm 0.0352$ \\
\bottomrule
\end{tabularx}
\end{table}

\begin{table}
\centering
\caption{Ablation study on the MONKEY dataset evaluating KongNet without ImageNet pretraining. Results are reported as mean FROC for inflammatory cells, lymphocytes, and monocytes across our internal 5-fold cross-validation on the MONKEY training set.}
\label{ablation_monkey_no_pretrain}
\begin{tabularx}{\linewidth}{
>{\raggedright\arraybackslash}p{2.9cm}
>{\centering\arraybackslash}X
>{\centering\arraybackslash}X
>{\centering\arraybackslash}X
}
\toprule
Method & Inflammatory Cells & Lymphocytes & Monocytes \\
\midrule
KongNet & \bm{$0.3537 \pm 0.0306$} & \bm{$0.3950 \pm 0.0378$} & \bm{$0.2407 \pm 0.0084$} \\
KongNet no pre-train & $0.3189 \pm 0.0391$ & $0.3578 \pm 0.0370$ & $0.1743 \pm 0.0333$ \\
\bottomrule
\end{tabularx}
\end{table}

\begin{table}
\centering
\caption{Summary of detection results (F1) in the MIDOG Challenge 2025 Track 1 \textbf{Preliminary Leaderboard}. \textsuperscript{1}YOLOv10~\cite{midog_cacfek}. \textsuperscript{2}RTMDet~\cite{midog_Gestalt}. \textsuperscript{3}YOLOv12~\cite{midog_CBIO}. \textsuperscript{4}SDF-YOLO~\cite{midog_YTU}.}
\label{midog_table_preliminary}
\begin{tabularx}{\columnwidth}{
>{\raggedright\arraybackslash}p{4cm}
>{\centering\arraybackslash}X
>{\centering\arraybackslash}X
>{\centering\arraybackslash}X
}
\toprule
Method & F1 $\downarrow$ & Precision & Recall \\
\midrule
KongNet-Det & \textbf{0.8147} & 0.7995 & 0.8306 \\
Team cacfek\textsuperscript{1} & 0.8127 & 0.7739 & \textbf{0.8556} \\
Team Gestalt\textsuperscript{2} & 0.8081 & 0.7818 & 0.8361 \\
Team CBIO\textsuperscript{3} & 0.8011 & \textbf{0.8079} & 0.7944 \\
Team YTUVisionRG\textsuperscript{4} & 0.7802 & 0.7539 & 0.8083 \\
\bottomrule
\end{tabularx}
\end{table}

\begin{table}
\centering
\caption{Ablation study on the MIDOG 2025 dataset comparing KongNet-Det trained using only the MIDOG 2025 training data with KongNet-Det trained using the MIDOG 2025 data together with additional external mitosis datasets. Results are reported as mean F1 score across our internal 5-fold cross-validation on the MIDOG 2025 training set.}
\label{ablation_midog_no_additional_datasets}
\begin{tabularx}{\linewidth}{
>{\raggedright\arraybackslash}p{8cm}
>{\centering\arraybackslash}X
}
\toprule
Method & F1 \\
\midrule
KongNet-Det without additional datasets & $0.7959 \pm 0.0120$ \\
KongNet-Det with additional datasets & \bm{$0.8314 \pm 0.0138$} \\
\bottomrule
\end{tabularx}
\end{table}

\begin{table}
\centering
\caption{=Ablation study on the MIDOG 2025 dataset comparing the full KongNet architecture, which uses auxiliary nuclei segmentation and contour segmentation tasks, with the detection-only KongNet-Det variant. Results are reported as mean F1 score across our internal 5-fold cross-validation on the MIDOG++ training set.}
\label{ablation_midog_full_kongnet}
\begin{tabularx}{\linewidth}{
>{\raggedright\arraybackslash}X
>{\centering\arraybackslash}X
}
\toprule
Method & F1 \\
\midrule
KongNet & \bm{$0.7967 \pm 0.0129$} \\
KongNet-Det & $0.7959 \pm 0.0120$ \\
\bottomrule
\end{tabularx}
\end{table}

\begin{table}[ht!]
\centering
\caption{Summary of detection results (F1) in the \textbf{PUMA Challenge Track 1} \textbf{Preliminary Leaderboard}. *Methods have not been publicly released at the time of writing. \textsuperscript{1}Multi-Stage Network~\cite{TeamLSM}. \textsuperscript{2}HoVer-NeXt model trained by challenge organisers~\cite{ref_puma_paper}. \textsuperscript{3}Team UME fine-tuned CellViT++ and performed a hyperparameter sweep~\cite{cellvit++puma}.}
\label{puma_table_track_1_preliminary}
\begin{tabularx}{\columnwidth}{
>{\raggedright\arraybackslash}p{4cm}
>{\centering\arraybackslash}X
>{\centering\arraybackslash}X
>{\centering\arraybackslash}X
>{\centering\arraybackslash}X
}
\toprule
Method & Average $\downarrow$ & Tumour & Lymphocytes & Other \\
\midrule
KongNet & \textbf{0.7060} & \textbf{0.8103} & \textbf{0.8070} & \textbf{0.5006} \\
Team LSM\textsuperscript{1} & 0.6779 & 0.7864 & 0.7938 & 0.4535 \\
Team rictoo* & 0.6761 & 0.7847 & 0.7714 & 0.4721 \\
Baseline\textsuperscript{2} & 0.6367 & 0.7337 & 0.7527 & 0.4235 \\
Team UME\textsuperscript{3} & 0.6109 & 0.7251 & 0.7435 & 0.3641 \\
\bottomrule
\end{tabularx}
\end{table}

\begin{table}[ht!]
\centering
\caption{Summary of detection results (F1) in the \textbf{PUMA Challenge Track 2} \textbf{Preliminary Test Set}. *Methods have not been publicly released at the time of writing. \textsuperscript{1}Multi-Stage Network~\cite{TeamLSM}. \textsuperscript{2}HoVer-NeXt model trained by challenge organisers~\cite{ref_puma_paper}. \textsuperscript{3}Team UME fine-tuned CellViT++ and performed a hyperparameter sweep~\cite{cellvit++puma}.}
\label{puma_table_track_2_preliminary}

\begin{subtable}{\textwidth}
\centering
\begin{tabularx}{\textwidth}{
>{\raggedright\arraybackslash}p{3cm}
*{6}{>{\centering\arraybackslash}X}
}
\toprule
Method & Average $\downarrow$ & Tum & Str & Apo & Epi & His \\
\midrule
KongNet & \textbf{0.2804} & \textbf{0.7981} & \textbf{0.4293} & 0.0286 & 0.08 & 0.2045 \\
Team LSM\textsuperscript{1} & 0.2760 & 0.7939 & 0.3398 & \textbf{0.1} & \textbf{0.0843} & \textbf{0.2965} \\
Team rictoo* & 0.2512 & 0.7896 & 0.3004 & 0.0286 & 0.0831 & 0.2834 \\
Baseline\textsuperscript{2} & 0.2274 & 0.7204 & 0.2238 & \textbf{0.1} & 0.025 & 0.1920 \\
Team UME\textsuperscript{3} & 0.2261 & 0.7094 & 0.2416 & 0.0472 & 0.08 & 0.2020 \\
\bottomrule
\end{tabularx}
\caption*{Part 1: Average; Tum: Tumour; Str: Stroma; Apo: Apoptosis; Epi: Epithelium; His: Histiocyte.}
\end{subtable}

\vspace{0.5em}

\begin{subtable}{\textwidth}
\centering
\begin{tabularx}{\textwidth}{
>{\raggedright\arraybackslash}p{3cm}
*{5}{>{\centering\arraybackslash}X}
}
\toprule
Method & Lym & Neu & End & Mel & Pla \\
\midrule
KongNet & \textbf{0.7877} & 0.0 & 0.1093 & \textbf{0.2418} & \textbf{0.1250} \\
Team LSM\textsuperscript{1} & 0.7444 & 0.0 & 0.1167 & 0.1719 & 0.1127 \\
Team rictoo* & 0.7099 & 0.0 & 0.1269 & 0.1896 & 0.0 \\
Baseline\textsuperscript{2} & 0.6923 & 0.0 & 0.0843 & 0.1635 & 0.0725 \\
Team UME\textsuperscript{3} & 0.7032 & 0.0 & \textbf{0.1818} & 0.2063 & 0.0530 \\
\bottomrule
\end{tabularx}
\caption*{Part 2: Lym: Lymphocyte; Neu: Neutrophil; End: Endothelium; Mel: Melanophage; Pla: Plasma.}
\end{subtable}

\end{table}

\begin{table}[ht!]
\centering
\caption{Summary of detection results (F1) on the \textbf{PanNuke Dataset} for each cell type and overall detection. \textbf{Ablation study on the effects of test-time augmentation}. $4\times$TTA provides the best balance between detection quality and computational efficiency. \textsuperscript{*}Class average (Class Avg) is calculated as the average F1 scores across five cell types. Neo: Neoplastic; Inf: Inflammatory; Epi: Epithelial; Con: Connective.}
\label{tta_ablation_pannuke_table}
\begin{tabularx}{\textwidth}{
>{\raggedright\arraybackslash}p{3.4cm}
>{\centering\arraybackslash}p{1cm}
>{\centering\arraybackslash}p{0.8cm}
>{\centering\arraybackslash}p{0.8cm}
>{\centering\arraybackslash}p{0.8cm}
>{\centering\arraybackslash}p{0.8cm}
>{\centering\arraybackslash}p{0.8cm}
>{\centering\arraybackslash}X
}
\toprule
Method & Overall & Neo & Inf & Epi & Con & Dead & Class Average\textsuperscript{*} \\
\midrule
KongNet $16\times$TTA & \textbf{0.84} & \textbf{0.71} & \textbf{0.72} & \textbf{0.65} & \textbf{0.70} & \textbf{0.59} & \textbf{0.674} \\
KongNet $4\times$TTA & 0.83 & 0.70 & 0.71 & 0.64 & \textbf{0.70} & \textbf{0.59} & 0.668 \\
KongNet No TTA & 0.81 & 0.68 & 0.69 & 0.61 & 0.67 & 0.56 & 0.642 \\
\bottomrule
\end{tabularx}
\end{table}

\begin{table}[ht!]
\centering
\caption{Inference time at $40\times$ magnification on the \textbf{workstation}. Processing time includes I/O and patch stitching. \textsuperscript{*}Models could not be evaluated due to memory or resource constraints. TTA: Test-time augmentation.}
\label{inference_time_40x_workstation}
\begin{tabularx}{\columnwidth}{
>{\raggedright\arraybackslash}p{6cm}
>{\centering\arraybackslash}X
>{\centering\arraybackslash}X
>{\centering\arraybackslash}X
}
\toprule
Method & Total Runtime (mins) & Inference Time (mins) & Processing Time (mins) \\
\midrule
KongNet no TTA & 8 & 6 & 2 \\
KongNet $4\times$ TTA & 26 & 24 & 2 \\
KongNet $16\times$ TTA & 94 & 92 & 2 \\
HoVer-NeXt Tiny $16\times$ TTA & 31 & 29 & 2 \\
CellViT-SAM H\textsuperscript{*} & * & * & * \\
\bottomrule
\end{tabularx}
\end{table}

\begin{table}[ht!]
\centering
\caption{Inference time at $40\times$ magnification on the \textbf{HPC node}. Processing time includes I/O and patch stitching. TTA: Test-time augmentation.}
\label{inference_time_40x_hpc}
\begin{tabularx}{\columnwidth}{
>{\raggedright\arraybackslash}p{6cm}
>{\centering\arraybackslash}X
>{\centering\arraybackslash}X
>{\centering\arraybackslash}X
}
\toprule
Method & Total Runtime (mins) & Inference Time (mins) & Processing Time (mins) \\
\midrule
KongNet no TTA & 6 & 3 & 3 \\
KongNet $4\times$ TTA & 14 & 11 & 3 \\
KongNet $16\times$ TTA & 43 & 40 & 3 \\
HoVer-NeXt Tiny $16\times$ TTA & 14 & 12 & 2 \\
CellViT-SAM H & 18 & 16 & 2 \\
\bottomrule
\end{tabularx}
\end{table}

\begin{table}[ht!]
\caption{Comparison of model forward pass time on the \textbf{workstation}.}\label{forward_pass}
\begin{tabularx}{\columnwidth}{
>{\raggedright\arraybackslash}X
>{\centering\arraybackslash}X
>{\centering\arraybackslash}X
}
\toprule
Method & Time (seconds) & Parameters \\
\midrule
% KongNet-Det & $0.0161 \pm 0.0010$ & 126 million \\
KongNet & $0.0378 \pm 0.0060$ & 176 million \\
HoVer-NeXt Tiny & $0.0133 \pm 0.0008$ & 54 million \\
CellViT-SAM-H & $0.1510 \pm 0.0009$ & 699 million \\
CellViT 256 & $0.0384 \pm 0.0008$ & 46 million \\
\bottomrule
\end{tabularx}
\end{table}

\begin{table}
\centering
\caption{Pairwise Davies--Bouldin (DB) index for MONKEY nuclei classes using hand-crafted features and ResNet18 deep features. Lower DB values indicate better cluster separation.}
\label{monkey_DB_index}
\begin{tabularx}{\linewidth}{
>{\centering\arraybackslash}X
>{\centering\arraybackslash}X
>{\centering\arraybackslash}X
>{\centering\arraybackslash}X
}
\toprule
Class 1 & Class 2 & Hand-crafted Features & ResNet18 Features \\
\midrule
Lymphocytes & Monocytes & 6.74 & 7.57 \\
\bottomrule
\end{tabularx}
\end{table}

\begin{table}
\centering
\caption{Pairwise Davies--Bouldin (DB) index for PanNuke nuclei classes using hand-crafted features and ResNet18 deep features. Lower DB values indicate better cluster separation.}
\label{pannuke_DB_index}
\begin{tabularx}{\linewidth}{
>{\centering\arraybackslash}X
>{\centering\arraybackslash}X
>{\centering\arraybackslash}X
>{\centering\arraybackslash}X
}
\toprule
Class 1 & Class 2 & Hand-crafted Features & ResNet18 Features \\
\midrule
Dead & Epithelial & 2.05 & 2.94 \\
Dead & Neoplastic & 1.80 & 2.73 \\
Dead & Inflammatory & 3.57 & 3.35 \\
Dead & Connective & 2.16 & 2.81 \\
Epithelial & Inflammatory & 3.84 & 6.76 \\
Epithelial & Connective & 5.08 & 7.08 \\
Epithelial & Neoplastic & 11.61 & 8.32 \\
Inflammatory & Connective & 2.97 & 4.74 \\
Inflammatory & Neoplastic & 2.98 & 4.91 \\
Connective & Neoplastic & 4.40 & 7.54 \\
\bottomrule
\end{tabularx}
\end{table}

\begin{table}
\centering
\caption{Ablation study on the PanNuke dataset evaluating automatic class-wise loss weighting ($\lambda$) for KongNet. Results are reported as F1 scores for each cell type, together with the overall and class-average F1. Neo: Neoplastic; Inf: Inflammatory; Epi: Epithelial; Con: Connective. Class average: the average of the F1 scores across five cell types.}
\label{ablation_pannuke_auto_lambda}
\begin{tabularx}{\textwidth}{
>{\raggedright\arraybackslash}p{3cm}
>{\centering\arraybackslash}p{1cm}
>{\centering\arraybackslash}p{0.8cm}
>{\centering\arraybackslash}p{0.8cm}
>{\centering\arraybackslash}p{0.8cm}
>{\centering\arraybackslash}p{0.8cm}
>{\centering\arraybackslash}p{0.8cm}
>{\centering\arraybackslash}X
}
\toprule
Method & Overall & Neo & Inf & Epi & Con & Dead & Class Average \\
\midrule
KongNet & \textbf{0.84} & \textbf{0.71} & \textbf{0.72} & \textbf{0.65} & \textbf{0.70} & \textbf{0.59} & \textbf{0.674} \\
KongNet auto $\lambda$ & 0.83 & 0.69 & 0.68 & 0.64 & \textbf{0.70} & 0.57 & 0.654 \\
\bottomrule
\end{tabularx}
\end{table}

\clearpage

\section{Supplementary Figures}

\begin{figure}[t!]
\centering
\includegraphics[width=0.25\linewidth]{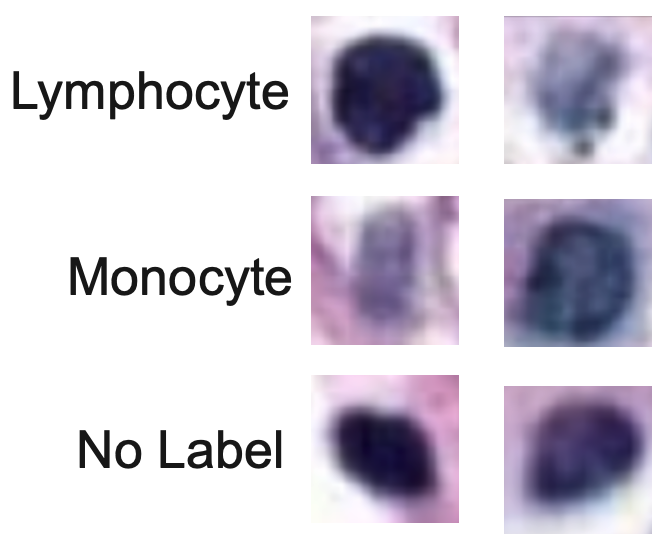}
\caption{Lymphocytes and monocytes in PAS-stained images from the MONKEY dataset. The subtle morphological differences between them make accurate visual distinction challenging, especially when cytoplasm is not stained.}
\label{monkey_lymph_mono}
\end{figure}

\begin{figure}[h!]
    \centering
    \includegraphics[width=\linewidth]{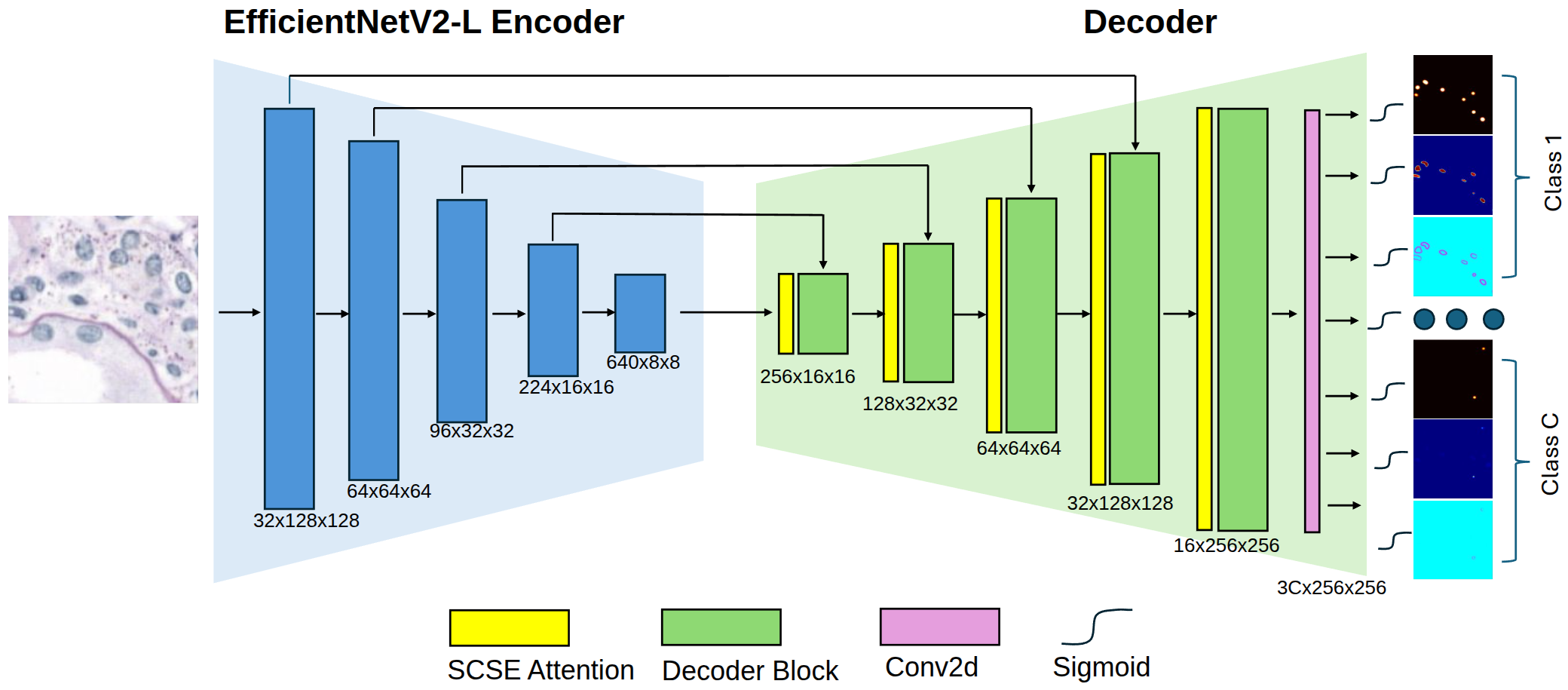}
    \caption{Architecture of KongNet-SH. The EfficientNetv2-L encoder is connected to one unified Decoder. The decoder predicts centroid map, segmentation map,  and contour map, for each cell class.}
    \label{KongNet-SH_vis}
\end{figure}

Visualization of KongNet-SH architecture is shown in Figure~\ref{KongNet-SH_vis}.

\begin{figure}[h!]
    \centering
    \includegraphics[width=\linewidth]{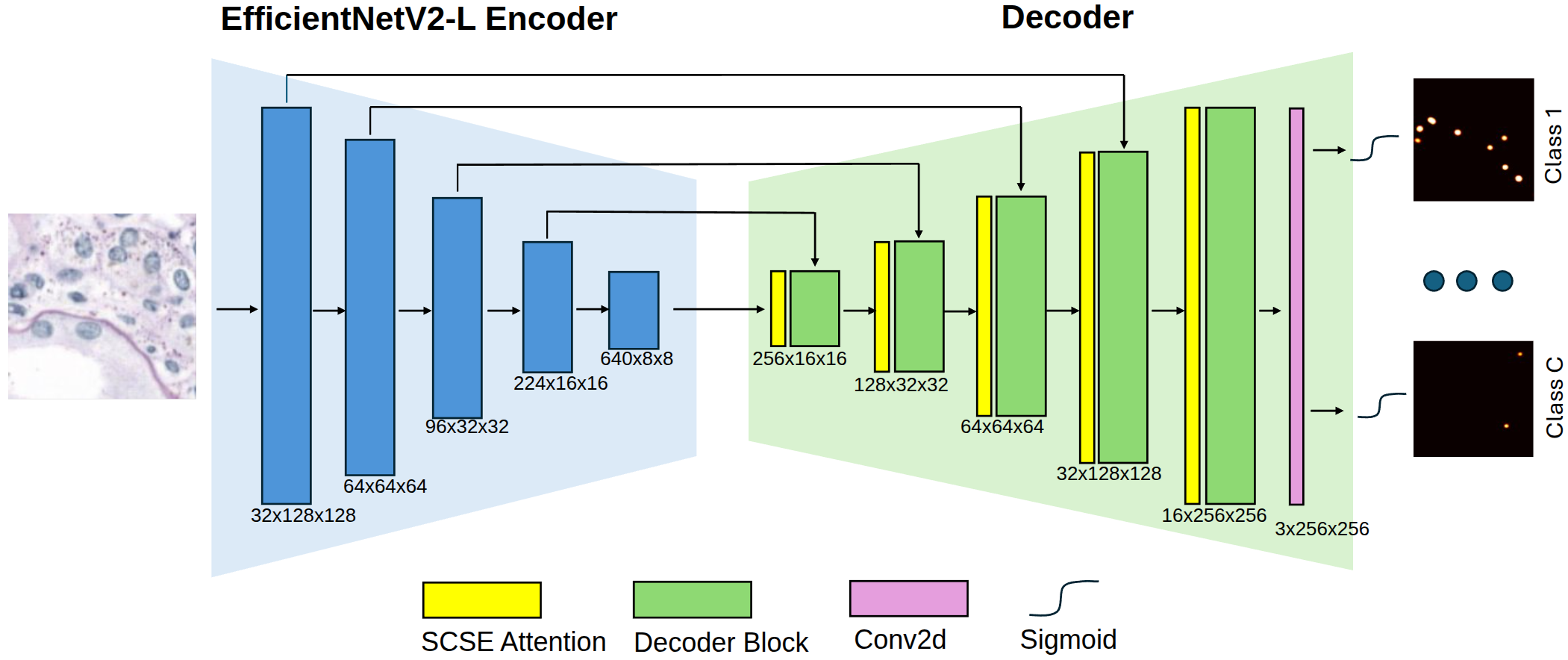}
    \caption{Architecture of KongNet-Det. The EfficientNetv2-L encoder is connected to one Decoder. The decoder predicts only centroid maps for each cell type.}
    \label{KongNet-Det_vis}
\end{figure}

Visualization of KongNet-Det architecture is shown in Figure~\ref{KongNet-Det_vis}.

\begin{figure}[h!]
    \centering
    \includegraphics[width=\linewidth]{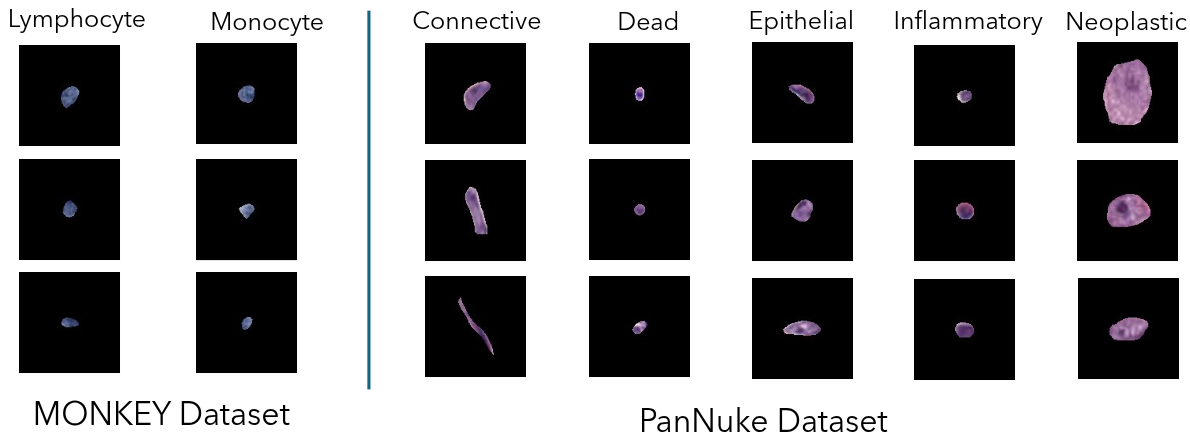}
    \caption{Visualisation of randomly selected nuclei instances from the MONKEY dataset and the PanNuke dataset.}
    \label{monkey_pannuke_nuclei}
\end{figure}

\begin{figure}[h!]
    \centering
    \includegraphics[width=\linewidth]{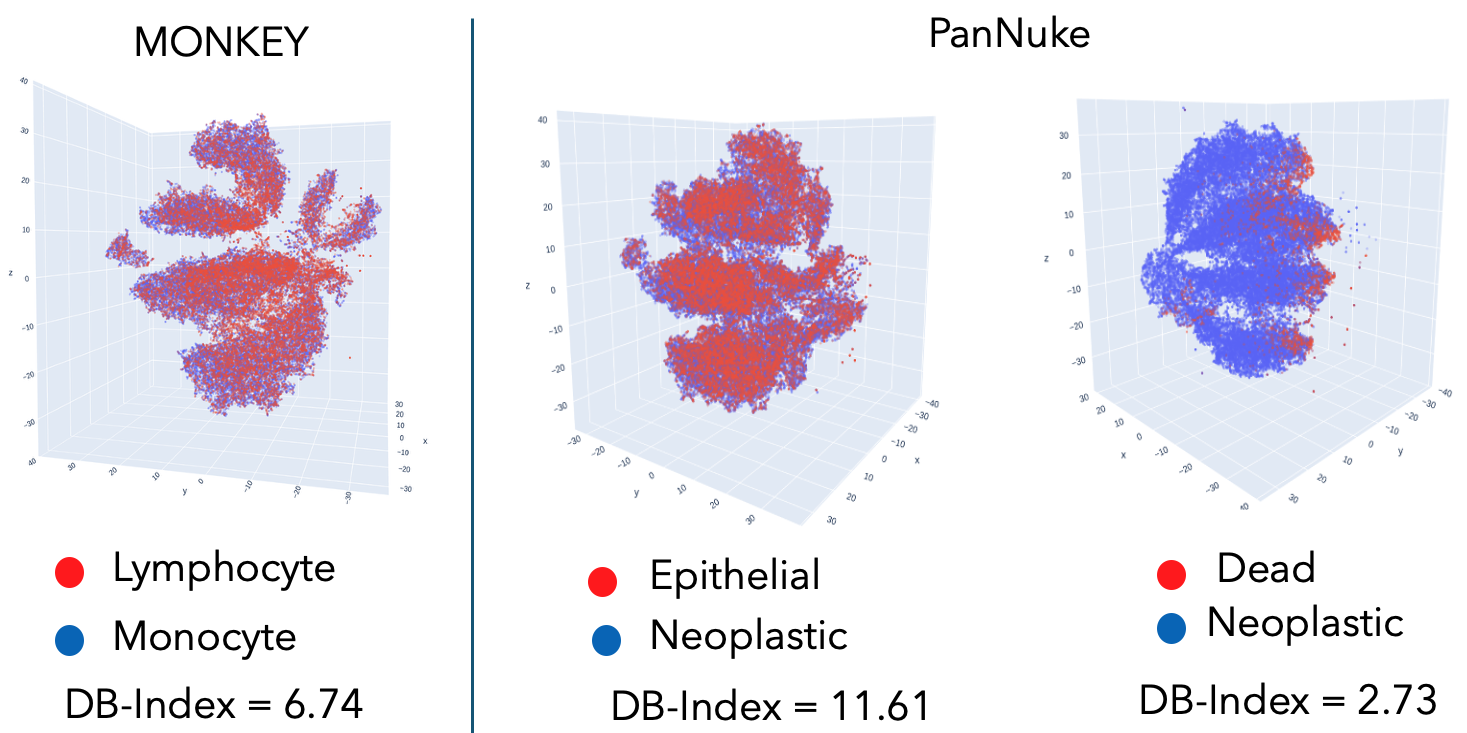}
    \caption{t-SNE visualisation of nuclei shape feature distributions for the MONKEY dataset and selected PanNuke class pairs. MONKEY lymphocytes and monocytes show substantial overlap, whereas in PanNuke the neoplastic and epithelial classes appear less well separated than the dead and neoplastic classes. This qualitative pattern is consistent with the corresponding Davies--Bouldin (DB) indices.}
    \label{tsne_plot}
\end{figure}

\end{document}